\documentclass[final,twocolumn]{IEEEtran} %draftcls, onecolumn

\usepackage{amssymb}          %%questo per usare mathbb
\usepackage{graphicx}         %%questo per la grafica
\usepackage[latin1]{inputenc} %%questo per scrivere lettere accentate direttamente da tastiera
\usepackage{amsmath}          %%questo per usare numerazioni avanzate

\usepackage{amsfonts}
\usepackage{latexsym}

\newtheorem{prop}{Proposition}
\newtheorem{defin}{Definition}
\newtheorem{thm}{Theorem}
\newtheorem{cor}{Corollary}
\newtheorem{lemma}{Lemma}

\newcommand{\proofn}{\noindent {\em Proof. }}

\newcommand{\vv}[1]{``#1''}

\newcommand{\ket}[1]{|#1\rangle}
\newcommand{\bra}[1]{\langle #1|}
\newcommand{\braket}[2]{\langle #1|#2\rangle}

\newcommand{\Hi}{\mathcal{H}}

\newcommand{\C}{\mathbb{C}}

\newcommand{\R}{\mathbb{R}}

\newcommand{\Bi}{\mathcal{B}}

\newcommand{\ddt}{\frac{d}{d t}}

\newcommand{\diag}{\textrm{diag}}

\newcommand{\tr}{\textrm{tr}}
\newcommand{\trace}{\textrm{trace}}

\def\smallfrac#1#2{{\textstyle\frac{#1}{#2}}}

\newcommand{\beq}{\begin{equation}}
\newcommand{\eeq}{\end{equation}}
\newcommand{\beqa}{\begin{eqnarray}}
\newcommand{\eeqa}{\end{eqnarray}}
\newcommand{\beqan}{\begin{eqnarray*}}
\newcommand{\eeqan}{\end{eqnarray*}}
\newcommand{\bea}{\begin{eqnarray}}
\newcommand{\eea}{\end{eqnarray}}

\title{Quantum Markovian Subsystems:\\ Invariance, Attractivity, and
Control}

\author{Francesco Ticozzi\thanks{F. Ticozzi is with the Dipartimento
di Ingegneria dell'Informazione, Universit\`a di Padova, via Gradenigo
6/B, 35131 Padova, Italy ({\tt ticozzi@dei.unipd.it}).} and Lorenza
Viola\thanks{L. Viola is with the Department of Physics and Astronomy,
Dartmouth College, 6127 Wilder Laboratory, Hanover, NH 03755, USA
({\tt Lorenza.Viola@Dartmouth.edu}).} }

\date{\today}

\begin{document}
\maketitle

\begin{abstract} 
We characterize the dynamical behavior of continuous-time, Markovian
quantum systems with respect to a subsystem of interest. Markovian
dynamics 
%accurately 
describes a wide class of open quantum systems of
relevance to quantum information processing, subsystem encodings
offering a general pathway to faithfully represent quantum
information.  We provide explicit linear-algebraic characterizations
of the notion of invariant and noiseless subsystem for Markovian
master equations, under different robustness assumptions for model-parameter and initial-state variations.  The stronger concept of an
attractive quantum subsystem is introduced, and sufficient existence
conditions are identified based on Lyapunov's stability techniques.
As a main control application, we address the potential of
output-feedback Markovian control strategies for quantum pure
state-stabilization and noiseless-subspace generation.  In particular,
explicit results for the synthesis of stabilizing semigroups and
noiseless subspaces in finite-dimensional Markovian systems are
obtained.
\end{abstract}

\vspace*{-2mm} 

\section{Introduction and Preliminaries}

Quantum subsystems are the basic building block for describing
composite systems in quantum mechanics~\cite{sakurai}.  From both a
conceptual and practical standpoint, renewed interest toward
characterizing quantum subsystems in a variety of control-theoretic
settings is motivated by Quantum Information Processing (QIP)
applications~\cite{nielsen-chuang}.  In order for abstractly defined
QIP protocols to be useful, information needs to be represented by
states of a physical system, in ways which minimize the impact of
errors and decoherence due to the interaction of the system with its
surrounding environment.  {\em Subsystem encodings} provide the most
general mathematical structure for realizing quantum information in
terms of physical degrees of freedom, and a main tool for achieving a
unified understanding of strategies for quantum error control in {\em
open} quantum systems~\cite{viola-generalnoise,knill-protected}.  In
particular, the idea that noise-protected subsystems may be identified
in the overall state space of a noisy physical system under
appropriate symmetry conditions underlies the method of passive
quantum stabilization via {\em decoherence-free subspaces}
(DFSs)~\cite{zanardi-DFS,lidar-DFS} and {\em noiseless subsystems}
(NSs)~\cite{viola-generalnoise,viola-qubit,kempe-NS}. In situations
where no such symmetry exists, open-loop dynamical decoupling
techniques can still ensure active protection through dynamical NS
synthesis~\cite{violaDygen} in the {\em non-Markovian} regime.

While substantial progress has been made toward defining and
exploiting subsystems within general quantum error correction theories
(see also Section II-A), the subsystem notion and its implications
have not yet reached out to the quantum control
community at large.  It is one of the goals of this paper to take a
step in this direction, by offering a {\em linear-algebraic}
characterization of the subsystem idea, which allows easier contact
with standard system-theoretic concepts such as invariance and
stability.  We focus on {\em continuous-time Markovian} quantum
dynamics \cite{alicki-lendi}, which both accurately describes a wide
class of open quantum systems and presents distinctive quantum
stabilization challenges compared to its non-Markovian counterpart.
In the process, we elucidate the key role played by different {\em
model robustness} properties in ensuring that desired dynamical
features may emerge, and in influencing the interplay between
Hamiltonian and dissipative components.  Beside leading to a
streamlined derivation of Markovian DFS- and NS- conditions which have
only partially appeared in the
literature~\cite{kempe-NS,lidar-initializationDFS}, our analysis
motivates the concept of {\em attractive
subsystem} as a strategy for ``dissipation-assisted'' asymptotic
initialization in an intended subsystem.  As a main application of our
work, we begin exploring the potential of Markovian, output-feedback
techniques for the robust synthesis of pure states and NSs in
finite-dimensional systems -- complementing the above-mentioned
open-loop dynamical schemes~\cite{violaDygen} as well as closed-loop
feedback approaches using continuous-time state
estimation~\cite{doherty-linear}.

\vspace{-2mm}

\vspace*{-2mm} 

\subsection{Notations} 

Consider a separable Hilbert space $\Hi$ over the complex field $\C$.
${\mathfrak B}(\Hi)$ represents the set of linear bounded operators on
$\Hi$, ${\mathfrak H}(\Hi)$ denoting the real subspace of Hermitian
operators, with ${\mathbb I}$, ${\mathbb O}$ being the identity and
the zero operator, respectively.  We indicate with $A^\dag$ the adjoint of
$A\in {\mathfrak B}(\Hi)$, with $c^*$ the conjugate of $c\in \C$. The
{\em commutator} and {\em anti-commutator} of $X,Y \in {\mathfrak
B}(\Hi)$ are defined as $[X,\,Y]:=XY-YX$, and $\{X,\,Y\}:=XY+YX$,
respectively.  In the special case where ${\cal H}$ is
two-dimensional, a convenient operator basis for the traceless sector
of ${\mathfrak B}(\Hi)$ is given by the Pauli operators,
$\sigma_\alpha, \alpha=x,y,z$, satisfying $[\sigma_\alpha,
\sigma_\beta]=2i \varepsilon_{\alpha\beta\gamma} \sigma_\gamma$, $\{
\sigma_\alpha, \sigma_\beta\} =2 \delta_{\alpha\beta}{\mathbb I}$,
$\varepsilon, \delta$ denoting the completely antisymmetric tensor and
the Kronecker delta, respectively.  
We choose the standard
representation where
$\sigma_x=
\left(
\begin{smallmatrix}
 0 & 1    \\
 1 &  0     
\end{smallmatrix}
\right),\,
\sigma_y=\left(
\begin{smallmatrix}
 0 & -i    \\
 i &  0     
\end{smallmatrix}
\right),\,
\sigma_z=\left(
\begin{smallmatrix}
 1 & 0    \\
 0 &  -1     
\end{smallmatrix}
\right).$

Throughout the paper we shall use the Dirac notation \cite{sakurai}.
Given the inner product $\langle\,,\,\rangle$ in ${\cal H}$, a natural
isomorphism exists between vectors in $\Hi$ (denoted $\ket{\psi}$, and
called a {\em ket}) and linear functionals in the dual space
$\Hi^\ast$ (denoted $\bra{\psi}$, and called a {\em bra}), so that
$\braket{\psi}{\varphi}=\langle\psi,\varphi\rangle$. If $A \in
{\mathfrak B}(\Hi)$, letting $A\ket{\psi}:=A(\psi)$ and
$\bra{\psi}A:=\langle A^\dag(\psi),\cdot\rangle$ yields
$\bra{\psi}A\ket{\varphi}=\langle\phi,A(\varphi)\rangle=\langle
A^\dag(\phi),\varphi\rangle$.  The \vv{outer} product
$\ket{\psi}\bra{\varphi}$ stands for $\langle \varphi, \cdot\rangle
\psi \in {\mathfrak B}(\Hi)$. Moreover, if $|\braket{\psi}{\psi}|=1,$
$\ket{\psi}\bra{\psi}=\langle \psi, \cdot\rangle \psi$ is the
orthogonal projector onto the one-dimensional subspace spanned by
$\ket{\psi}$.

\vspace*{-2mm} 

\subsection{Basic notions of statistical quantum mechanics} 

In the standard formulation of quantum mechanics \cite{sakurai}, a
quantum system ${\cal Q}$ is associated with a separable Hilbert space
$\Hi$, whose dimension is determined by the physics of the problem.
{\em Physical observables} are modeled as self-adjoint operators on
$\Hi$, the set of possible outcomes they can assume in a von Neumann
measurement process being their spectrum. In what follows, we consider
only observables with {\em finite} spectrum, that can be represented
as Hermitian matrices acting on $\Hi \simeq \C^d, d < \infty$.  Our
(possibly uncertain) knowledge of the state of $\mathcal{Q}$ is
condensed in a {\em density operator} $\rho$, with $\rho\geq 0$ and
$\trace(\rho)=1$.  Density operators form a convex set ${\mathfrak
D}(\Hi)\subset {\mathfrak H}(\Hi)$, with one-dimensional projectors
corresponding to extreme points ({\em pure states},
$\rho_{|\psi\rangle}=|\psi\rangle\langle \psi|$).

If ${\cal Q}$ comprises two quantum {\em subsystems} ${\cal
Q}_1,\,{\cal Q}_2$, the corresponding mathematical description is
carried out in the tensor product space, $\Hi_{12}=\Hi_1\otimes\Hi_2$
\cite{sakurai}, observables and density operators being associated
with Hermitian, positive-semidefinite, normalized operators on
$\Hi_{12}$, respectively.  
\iffalse
A density operator $\rho_{12} \in
{\mathfrak D}({\cal H}_{12})$ is {\em separable} if it may be
expressed as a convex sum of product states, that is,
$\rho_{12}=\sum_a w_a \rho_1^a \otimes \rho_2^a$, with $\rho_i^a \in
{\mathfrak D}(\Hi_i)$. 
\fi
In particular, a joint pure state $\rho_{12}=|\psi\rangle_{12}\langle
\psi|$ which cannot be factorized into the product of two pure states
on the individual factors is called {\em entangled}.  Let $X$ be an
element of ${\mathfrak B}(\Hi_{12})$.  The {\em partial trace} over
$\Hi_2$ is the unique linear operator $\trace_{2}(\cdot): {\mathfrak
B}(\Hi_{12})\rightarrow {\mathfrak B}(\Hi_{1})$, ensuring that for
every $X_1\in{\mathfrak B}(\Hi_{1}),X_2\in{\mathfrak B}(\Hi_{2})$,
$\trace_{2}(X_1\otimes X_2)=X_1 \trace(X_2)$.  If $\rho_{12}$
represents the joint density operator of ${\cal Q}$, the state of,
say, subsystem ${\cal Q}_1$ alone is described by the {\em reduced
density operator} $\rho_1=\trace_{2}(\rho_{12}) \in {\mathfrak
H}(\Hi_{1})$, so that $\trace(\rho_1 X_1) = \trace(\rho_{12}
X_1\otimes {\mathbb I}_2)$, for all observables $A_1\in {\mathfrak
H}(\Hi_{1})$.

We consider quantum dynamics in the Schr\"odinger picture, with pure
states and density operators evolving forward in time, and
time-invariant observables.  The evolution of an {\em isolated}
(closed) quantum system is driven by the Hamiltonian, $H$, according
to the Liouville-von Neumann equation:
\[
\label{eq:liouvillian} \frac{d}{dt}\rho(t)=-i[H,\rho(t)]\,,
\]
\noindent
in units where $\hbar=1$.  Thus, $\rho(t)=U_t\rho(0)U_t^\dag,$ where
the unitary evolution operator (or {\em propagator}) $U_t=e^{-iHt}$.

In general, in the presence of internal couplings, quantum
measurements, or interaction with a surrounding environment, the
evolution of a subsystem of interest is no longer unitary and
reversible, and the general formalism of {\em open quantum systems} is
required \cite{davies,alicki-lendi,petruccione}.  The physically
admissible discrete-time evolutions for a quantum system may be
characterized axiomatically and are called {\em quantum operations},
or completely positive (CP) maps \cite{kraus}.  Let ${\cal I}$ denote
the physical quantum system of interest, with associated Hilbert space
$\Hi_I$, $\textrm{dim}(\Hi_I)=d$. The class of trace-preserving (TP)
quantum operations is relevant to our purposes:

\begin{defin}[TPCP map] 
A TPCP-map $\mathcal{T}(\cdot)$ on ${\cal I}$ is a map on
${\mathfrak D}({\cal H}_I)$, that satisfies:
\begin{itemize}
\item [(i)] $\mathcal{T}(\cdot)$ is {convex linear}: given states
$\rho_{i}\in {\mathfrak D}({\cal H}_{I}),$
$$\mathcal{T}\Big(\sum_i p_i \rho_i\Big)=\sum_i p_i
\mathcal{T}(\rho_i),\quad\sum_i p_i=1,\, p_i \geq 0, \, \forall i;$$
\item [(ii)] $\mathcal{T}(\cdot)$ is trace-preserving:
$$ \trace(\mathcal{T}(\rho_I))= \trace(\rho_I)=1; $$
\item [(iii)] $\mathcal{T}(\cdot)$ is {completely positive}:
$$\mathcal{T}(\rho_I)\geq0,\quad ({\mathbb
I}_m\otimes\mathcal{T})(\rho_{I E})\geq0, $$
\noindent
for every $m$-dimensional ancillary space $\Hi_E$ joint to $\Hi_I$,
and for every $\rho_{I E}\in {\mathfrak D}({\cal H}_{I}\otimes{\cal
H}_{E}) $~\footnote{The CP assumption is necessary to preserve
positivity of arbitrary purifications of states on ${\cal H}_I$,
including entangled states, see e.g. \cite{nielsen-chuang} for a
discussion of a well-known counter-example, based on the transpose
operation.}.
\end{itemize}
\end{defin}

\vspace*{1mm}

A more concrete characterization of the dynamical maps of interest is
provided by the following:

\begin{thm}[Hellwig-Kraus representation theorem] $\mathcal{T}[\cdot]$ 
is a TPCP map on ${\cal I}$ iff for every $\rho_I\in{\mathfrak
D}(\Hi_I)$:
$$\mathcal{T}(\rho_I)=\sum_k E_k\rho_I E_k^\dag, $$
\noindent 
where $\{E_k\}$ is a family of operators in ${\frak B}(\Hi_I)$ such
that $\sum_kE_k^\dag E_k = {\mathbb I}$. \end{thm}

\vspace*{1mm} 

As a consequence of the above Theorem, every TPCP map
$\mathcal{T}(\cdot)$ on ${\mathfrak D}(\Hi_I)$ may be extended to a
well-defined linear, positive, and TP map on the whole ${\mathfrak
B}(\Hi_I)$.

\vspace*{-2mm} 

\subsection{Quantum dynamical semigroups}

A relevant class of open quantum systems obeys Markovian dynamics
\cite{alicki-lendi,lindblad,qds}. Assume that we have no access to the
quantum environment surrounding ${\cal I}$, and that the dynamics in
${\mathfrak D}(\Hi_I)$ is continuous in time, the state change at each
$t >0$ being described by a TPCP map ${\cal T}_t(\cdot)$.  A
differential equation for the density operator of ${\cal I}$ may be
derived provided that a forward composition law holds:

\begin{defin}[QDS]\label{QDS} 
A {\em quantum dynamical semigroup} is a one-parameter family of TPCP
maps $\{{\cal T}_t(\cdot),\,t\geq0\}$ that satisfies:
\begin{itemize}
\item [(i)] ${\cal T}_0={\mathbb I}$, 
\item [(ii)] ${\cal T}_t\circ{\cal T}_s={\cal T}_{t+s},\; \forall t,s >0,$  
\item [(iii)] $\textrm{{trace}}({\cal T}_t(\rho) X)$ is a continuous
function of $t$, $\forall\rho\in{\mathfrak D}(\Hi_I)$, $\forall X\in
\Bi(\Hi_I)$.
\end{itemize}
\end{defin}

\vspace*{1mm} 

\noindent 
Due to the trace and positivity preserving assumptions, a QDS is a
semigroup of contractions\footnote{A map $f$ from a metric space
$\mathfrak{M}$ with distance $d(\cdot, \cdot)$ to itself is a {\em
contraction} if $d(f(X),f(Y))\leq kd(X,Y)$ for all $X,Y$ in
$\mathfrak{M}$, $0<k\leq 1$.}.  It has been proved
\cite{lindblad,gorini-k-s} that the Hille-Yoshida generator for a QDS
\cite{hille-phillips} exists and can be cast in the following
canonical form:
\begin{eqnarray}
\hspace*{-10mm} 
&&\ddt
\rho(t)=\mathcal{L}(\rho(t))=-{i}[H,\rho(t)]+\sum_{k,l=1}^{d^2-1}{\cal L}_{kl}(\rho(t))
\label{eq:GKS}\\
\hspace*{-5mm}
&&=-{i}[H,\rho(t)]+\frac{1}{2} \sum_{k,l=1}^{d^2-1}
a_{kl}\left(2F_k\rho(t)F_l^\dag-\{F_k^\dag F_l,\,\rho(t)\}\right),
\nonumber
\end{eqnarray}
where $\{F_k\}^{d^2-1}_{k=0}$ is a basis of ${\mathfrak B} (\Hi_I),$
the space of linear operators on $\Hi_I$, with $F_0={\mathbb I}$. The
positive definite $(d^2-1)$-dimensional matrix $A=(a_{kl})$, which
physically specifies the relevant relaxation parameters, is also
called the Gorini-Kossakowski-Sudarshan (GKS) matrix.  Thanks to the
Hermitian character of $A$, Eq. \eqref{eq:GKS} can be rewritten in a
symmetrized form, moving to an operator basis that diagonalizes $A$:
\begin{eqnarray}
\hspace*{-10mm}&&%\ddt \rho(t)%=
\mathcal{L}(\rho(t))=
-{i}
[H,\rho(t)]+\sum_{k=1}^{d^2-1}\gamma_k {\cal
D}(L_k,\rho(t))\,\label{eq:lindblad}\\
\hspace*{-7mm}&&
=-{i}[H,\rho(t)]+\sum_{k=1}^{d^2-1}
{\gamma_k}\Big(L_k\rho(t)L_k^\dag-\frac{1}{2}\{L_k^\dag
L_k,\,\rho(t)\}\Big), \nonumber
\end{eqnarray}
with $\{\gamma_k\}$ denoting the spectrum of $A$. The {\em effective
Hamiltonian} $H$ and the {\em Lindblad operators} $L_k$ specify the
net effect of the Markovian environment on the dynamics.  In general,
$H$ is equal to the isolated system Hamiltonian, $H_0$, plus a
correction, $H_L$, induced by the coupling to the environment
(so-called Lamb shift).  The non-Hamiltonian terms ${\cal
D}(L_k,\rho(t))$ in \eqref{eq:lindblad} account for non-unitary
dynamics induced by $L_k$.

\vspace*{-2mm} 

\subsection{Phenomenological Markovian models and robustness}
\label{sec:robustness}

In principle, the exact form of the generator of a QDS may be
rigorously derived from the underlying Hamiltonian model for the joint
system-environment dynamics under appropriate limiting conditions (the
so-called ``singular coupling limit'' or the ``weak coupling limit,''
respectively~\cite{alicki-lendi,petruccione}).  In most physical
situations, however, carrying out such a procedure is unfeasible,
typically due to lack of complete knowledge of the full microscopic
Hamiltonian.  A Markovian generator of the form \eqref{eq:GKS} is then
assumed on a phenomenological basis\footnote{The Markovian generator
can be inferred from experimentally available data via quantum process
tomography, see e.g. \cite{nielsen-chuang}. By measuring the effect of
the environment on known states for fixed times, one may reconstruct
the corresponding set of CP maps and the underlying infinitesimal
generator.  Special care must be paid to the fact that this procedure
may lead to non-CP maps in the presence of measurement or numerical
errors \cite{Boulant}.  In some situations, approximate models may be
obtained upon formal quantization of a classical master equation,
e.g. the so-called Pauli master equation \cite{alicki-lendi}.}. In
practice, it is often the case that direct knowledge of the noise
effect is available, allowing one to specify the Markovian generator
by either giving a GKS matrix in \eqref{eq:GKS} or a set of noise
strengths $\gamma_k$ and Lindblad operators $L_k$ (not necessarily
orthogonal or complete) in \eqref{eq:lindblad}.  Each of the noise
operators $L_k$ may be thought of as corresponding to a distinct {\em
noise channel} ${\cal D}(L_k,\rho(t))$, by which information
irreversibly leaks from the system into the environment.

\vspace*{1mm} 

{\em Example 1:} Consider a two-level atom, with ground and excited
states $\ket{\psi_g},\,\ket{\psi_e}$, respectively. Assume that there
is an average rate of decay from the excited to the ground state
$\gamma >0$, that is, the survival probability of the excited state
$\ket{\psi_e}$ decays as $e^{-\gamma t}$. The resulting dynamics is
well described by a semigroup master equation of the form
%\beq\label{example1} 
$$\ddt \rho(t)=-{i}\omega
[\sigma_z,\rho(t)]+\frac{\gamma}{2}\Big( 2\sigma_- \rho(t)
\sigma_+ -\{\sigma_+ \sigma_- ,\,\rho(t)\}\Big), $$
\noindent
where $\omega >0$ determines the energy splitting between the ground
and excited state, and $\sigma_-=\ket{\psi_g}\bra{\psi_e}$, $\sigma_+=
\sigma_-^\dagger$ are pseudo-spin lowering and raising operators,
respectively.  In fact, computing the probability ${\mathbb
P}_e(t)=\trace(\rho(0)\rho(t)),$ with initial state
$\rho(0)=|\psi_e\rangle\langle \psi_e|$, yields ${\mathbb
P}_e(t)=e^{-\gamma t}$.

\vspace*{1mm} 

In the above example, a single noise channel, ${\cal
D}(\sigma_-,\rho(t))$, is relevant.  In physical situations, it often
happens that known properties of the error process naturally restrict
the relevant Lindblad operators or the admissible GKS descriptions to
a {\em reduced form} which incorporates the existing constraints.  A
paradigmatic QIP-motivated example is the following:

\vspace*{1mm} 

{\em Example 2:} Consider a {\em quantum register} ${\cal Q}$, that
is, a quantum system composed by $q$ two-dimensional systems ({\em
qubits}), with associated state $\Hi_Q =
\C^2_{(1)}\otimes\cdots\otimes \C^2_{(q)}$, $d=2^q$.  For an arbitrary
Markovian error process, combined errors on any subset of qubits may
occur, corresponding to an error basis $\{ F_k\}$ which spans the full
traceless sector of ${\mathfrak B} (\Hi_Q)$ or, equivalently, the
($d^2-1$)-dimensional Lie algebra $\mathfrak{su}(d)$.  Under the
assumption that {\em linear decoherence} takes place, errors can
independently affect at most one qubit at the time, reducing the
relevant error set to operators of the form
~\cite{lidar-DFS,viola-generalnoise}:
$$F_{k,l}=\mathbb{I}^{(1)}\otimes\cdots
\mathbb{I}^{(k-1)}\otimes\sigma^{(k)}_l\otimes
\mathbb{I}^{(k+1)}\otimes\cdots\otimes \mathbb{I}^{(q)},$$
\noindent
where $k=1,\ldots, q$, and $l=x,y,x$. Completing these $3q$
orthonormal generators to the above basis for the traceless operators
in ${\mathfrak B}(\Hi_Q)$, Eq.~\eqref{eq:GKS} formally holds. Clearly,
the resulting matrix $A$ differs from zero only in a $(3q\times
3q)$-dimensional block.  If the noise process is additionally
restricted to obey permutational symmetry (so-called {\em collective
decoherence}), the relevant error set is further reduced to completely
symmetric generators of the form
$$F_{l}= \sum_{k=1}^q
\mathbb{I}^{(1)}\otimes\cdots
\otimes\sigma^{(k)}_l\otimes
\cdots\otimes \mathbb{I}^{(q)},\;\;\;l=x,y,z,$$
\noindent
in which case $\mbox{span}\{ F_k\} \simeq \mathfrak{su}(2)$, and $A$
may effectively be taken as a $3\times 3$ positive-definite matrix.

\vspace*{1mm} 

The above examples show how, in practice, a compact version of
\eqref{eq:GKS} typically suffices, in terms of (orthonormal) error
generators $\{F_k\}$ spanning a $m$-dimensional {\em error
subalgebra}, $m \leq d^2-1$.  The corresponding Markovian generator in
\eqref{eq:GKS} is then completely specified by a {\em reduced GKS
matrix} of dimension $m \times m$.
 
In the following sections, we are interested at characterizing
dynamical properties of finite-dimensional QDSs in terms of their
generator.  As in most situations only limited or approximate
knowledge about the model is available, we are naturally led to
consider two kinds of {\em structured
robustness}~\cite{doyle-feedback,ticozzi-robust}:

\vspace*{1mm} 

\begin{defin}[Model robustness] \label{robustproperty} 
Assume that a system ${\cal I}$ undergoes QDS dynamics, under a
nominal generator of the form \eqref{eq:GKS} or \eqref{eq:lindblad},
with uncertain knowledge of the parameters $A=(a_{jk})_{j,k=1}^m$ or
$\Gamma =(\gamma_1,\ldots,\gamma_m)$.  Let ${\cal A}$ and ${\cal V }$
denote the uncertainty sets, that is, the sets of parameters
identifying the admissible models in the form \eqref{eq:GKS} and
\eqref{eq:lindblad}, respectively.  A property ${\mathfrak P}$ is said
to be
\begin{itemize}
\item[{ (i)}] {{\em $A$-robust}} if it holds for every $A=(a_{ij})\in
{\cal A}$;
\item[{ (ii)}] {{\em $\gamma$-robust}} if it holds for every
$\Gamma = (\gamma_1,\ldots,\gamma_m)\in {\cal V}$. 
\end{itemize}
\end{defin}

\vspace*{1mm}

\noindent 
The study of $A$- or $\gamma$-robustness is important to establish
whether a desired feature of the model (e.g. invariance of a
subsystem, existence of attractive states) may be ensured by avoiding
{\em fine-tuning} on the noise parameters
\cite{lidar-DFS}.

{\em Remarks:} $A$-robustness implies $\gamma$-robustness. However the
converse is not true. In fact, $\gamma$-robustness corresponds to
robustness only with respect to variation in the spectrum of $A$.
While studying $A$-robustness, we shall always imply a reduced
description as explained above, with ${\cal
A}=\{A=(a_{k,l})_{k,l=1}^m\}$ denoting the set of relevant reduced GKS
matrices. Clearly, only properties that are completely independent of
the particular noise model can be $A$-robust if ${\cal A}$ is the
whole ${\cal B}(\Hi_I).$

\vspace*{-2mm} 

\section{Theory}

\vspace*{-2mm} 

\subsection{Quantum subsystems and their role in QIP}

In order for the physical system ${\cal I}$ to implement a QIP task,
it is necessary that at every point in time well-defined ``logical''
degrees of freedom exist, which carry the desired quantum information
and support a basic set of control capabilities.  Within the standard
quantum network model~\cite{nielsen-chuang}, such set must include the
ability to:

\begin{itemize} 
\item {\bf Unitary control}: Implement a set of control actions that
ensure universal control. 
\item {\bf Initialization}: Realize a quantum operation that prepares
the system in an intended pure state.
\item {\bf Read-out}: Perform measurements of appropriate system's
observables\footnote{In the simplest setting, the ability to effect
strong, von Neumann projective measurements is assumed, which together
with unitary control implies the ability to initialize the state.}.
 
\end{itemize}

According to the {\em subsystem
principle}~\cite{viola-qubit,knill-protected}, the most general
structure which can faithfully embody quantum information is a
subsystem of ${\cal I}$.  Intuitively, a subsystem may be thought of
as a \vv{portion} of the full physical system, whose states, in the
simplest setting, obey {\em perfectly} the criteria above. Logical
subsystems may or may not directly coincide with physically natural
degrees of freedom.  If ${\cal I}$ is noisy, in particular, it
suffices that the action of noise be either negligible or correctable
on subsystem where the information resides.  Thus, protecting
information need not require the full state of the physical system to
be immune to noise, although this typically involves encodings which
are entangled with respect to the natural subsystem degrees of
freedom. A paradigmatic example is the protected qubit encoded in
three spin-$1/2$ particles subject to collective
decoherence~\cite{viola-generalnoise,viola-qubit}.

Formally, the following definition is suitable to QIP settings:

\vspace*{1mm}

\begin{defin}[Quantum subsystem] \label{def:subsys}
A {\em quantum subsystem} $\mathcal{S}$ of a system $\mathcal{I}$
defined on $\Hi_I$ is a quantum system whose state space is a tensor
factor $\Hi_S$ of a subspace $\Hi_{SF}$ of $\Hi_I$, %that is, 
\beq\Hi_I
=\Hi_{SF}\oplus \Hi_R= (\Hi_{S}\otimes\Hi_F)\oplus \Hi_R,
\label{eq:subs}\eeq
\noindent
for some factor ${\cal H}_F$ and remainder space ${\cal H}_R$.  The
set of linear operators on ${\cal S}$, ${\cal B}({\cal H}_S)$, is
isomorphic to the (associative) algebra on $\Hi_I$ of the form
$X_I=X_{S}\otimes \mathbb{I}_F \oplus {\mathbb O}_R$.
\end{defin}

\vspace*{1mm}

Let $n=\dim(\Hi_S),$ $f=\dim(\Hi_F),$ $r=\dim(\Hi_R)$, and let
$\{\ket{\phi_j^{S}}\}_{j=1}^n,\,\{\ket{\phi_k^{F}}\}_{k=1}^f,\,
\{\ket{\phi_l^{R}}\}_{l=1}^r$ denote orthonormal bases for
$\Hi_{S},\,\Hi_F,\,\Hi_R,$ respectively.  Decomposition
\eqref{eq:subs} is then naturally associated with the following basis
for $\Hi_I$:
$$\{\ket{\varphi_m}\}=\{\ket{\phi_j^{S}}
\otimes\ket{\phi_k^{F}}\}_{j,k=1}^{n,f}\cup\{\ket{\phi_l^{R}}\}_{l=1}^r.$$
\noindent
This basis induces a block structure for matrices acting on $\Hi_{I}$:
\beq\label{eq:blocks} X=\left(
\begin{array}{c|c}
  X_{SF} & X_P     \\\hline
  X_Q  &  X_R   
\end{array}
\right), \eeq 
\noindent
where, in general, $X_{SF}\neq X_{S}\otimes X_F$. Let $\Pi_{SF}$ be
the projection operator onto $\Hi_S\otimes\Hi_F$, that is,
$\Pi_{SF}=\left(\begin{array}{c|c} \mathbb{I}_{SF} &
0\end{array}\right)$.

For a noisy system ${\cal I}$, the goal of {\em passive} quantum error
control is to identify subsystems of ${\cal I}$ where the dominant
error events have minimum (ideally no) effect.  Loosely speaking, each
error operator belonging to the fixed error set for which protection
is sought must have an ``identity action'' once appropriately
restricted to the intended subsystem.  Historically, the first kind of
subsystems considered to this purpose have been {\em noiseless
subspaces} of the system's Hilbert space, often called DFSs in the
relevant literature~\cite{zanardi-DFS,lidar-DFS}:
$$\Hi_I=\Hi_{DFS}\oplus \Hi_R,$$ 
\noindent
which corresponds to a special instance of decomposition
\eqref{eq:subs} with one-dimensional ``syndrome'' co-subsystem ${\cal
F}$, $\Hi_F\simeq\mathbb{C}$.  The possibility for genuine {\em
noiseless subsystem}-encodings to exist and be useful was recognized
in~\cite{viola-generalnoise}, in which case we specialize the notation
of \eqref{eq:subs} to 
\beq\Hi_I=(\Hi_{NS}\otimes\Hi_F)\oplus
\Hi_R,\label{eq:H_SS}
\eeq
\noindent
by explicitly identifying the noiseless factor with ${\cal
H}_{NS}\,$\footnote{In principle, multiple NSs may exist for a given
dynamical system.  While we do not explicitly address such a scenario,
generalization is possible along the lines presented here.}.

DFS and NS theory has received extensive attention to date. A
relatively straightforward characterization is possible for error
sets which are effectively $\dagger$-closed, in which case elegant
results from the representation theory of C*-algebras apply\footnote{A
C*-algebra is a complex normed algebra $\mathcal{A}$ with a conjugate
linear involution (* or $\dagger$, an {\em adjoint} operation), which
is complete, satisfies $\|AB\|\leq\|A\|\|B\|,$ and $\|A^\dagger
A\|=\|A\|^2$, for all $A,B\in\mathcal{A}$. Any norm-closed subspace of
bounded operators on ${\mathcal H}$ is a C*-algebra if it closed under
the usual adjoint operation. Up to unitary equivalence, every
finite-dimensional operator *-algebra is isomorphic to a unique direct
sum of ampliated full matrix algebras.  Such a decomposition directly
reveals the supported NSs, whenever ${\mathcal A}$ represents an {\em
error algebra} for the noisy system ${\cal
I}$~\cite{viola-generalnoise}.}.  The operator-algebraic approach is
suitable for investigating NSs within both a Hamiltonian formulation
of open-system dynamics and a large class of TPCP maps, see
e.g.~\cite{viola-generalnoise,kribs-prl}.  However, explicit
characterizations for arbitrary quantum operations and Markovian
dynamics are more delicate and, to some extent, less consolidated.
While a number of definitions and results are provided
in~\cite{kempe-NS,lidar-initializationDFS,verification}, the
increasingly prominent role that quantum subsystems play within
quantum error correction theory \cite{Klappenecker07}, along with
continuous experimental advances in implementing DFSs
\cite{Kielpinski,violadfs,nicolasQEC} and NSs \cite{viola-science},
heighten the need for a fully consistent system-theoretic
approach.  It is our goal in the remaining of this Section to provide
such a framework for the case of Markovian dynamics, by paying special
attention to the key role played by model robustness notions as stated
in Definition 3.

\vspace*{-2mm} 

\subsection{Invariant subsystems}

%\begin{defin}[Perfect state initialization]\label{initialization} 
\begin{defin}[State initialization]\label{initialization} 
The system $\mathcal{I}$ with state $\rho\in{\mathfrak D}(\Hi_{I})$ is
{\em %perfectly 
initialized in $\Hi_{S}$ with state $\rho_{S}
\in{\mathfrak D}(\Hi_{S})$} if the blocks of $\rho$ satisfy:
\begin{itemize}   
\item[{(i)}] $\rho_{SF}=\rho_{S}\otimes\rho_F$ for some
$\rho_{F}\in{\mathfrak D}(\Hi_{F});$

\item[{(ii)}] $\rho_P=0,\rho_R=0.$
\end{itemize}
\end{defin}

%\vspace*{1mm}

Condition (ii) in the above Definition guarantees that
$\bar{\rho}_S=\trace_F(\Pi_{SF}\rho \Pi_{SF}^\dag)$ is a valid state
of ${\cal S}$, while condition (i) ensures that measurements or
dynamics affecting the factor $\Hi_F$ have no effect on the state in
$\Hi_S$. We shall denote by ${\mathfrak I}_S(\Hi_I)$ the set of states
initialized in this way.  The larger set of states obeying condition
(ii) alone will correspondingly be denoted by ${\mathfrak
I}_{SF}(\Hi_I)$.

\vspace*{1mm}

\begin{defin}[{Invariance}] \label{invariance} 
Let ${\cal I}$ evolve under TPCP maps.
${\cal S}$ is an {\em invariant subsystem} if 
the evolution of $\rho \in{\mathfrak I}_{S}(\Hi_I)$ obeys:
\begin{equation}\label{eq:invariance}
\rho(t)=\left(
\begin{array}{c|c}
  {\cal T}^S_t(\rho_{S})\otimes{\cal T}^F_t(\rho_F) & 0    \\\hline
  0  &  0  
\end{array}
\right),\;\; t\geq0,
\end{equation}
$\forall\,\rho_{S}\in{\mathfrak D}(\Hi_{S}),\,\rho_{F}\in{\mathfrak D}
(\Hi_{F})$, and with ${\cal T}^S_t(\cdot)$ and ${\cal T}^F_t(\cdot)$, 
$t\geq 0$, being TPCP maps on $\Hi_S$ and $\Hi_F$, respectively.
\end{defin}

\vspace*{1mm}

Thus, a subsystem is invariant if time evolution preserves the %perfect
initialization of the state, that is, the dynamics is confined within
${\mathfrak I}_S(\Hi_I)$. For Markovian evolution of $\mathcal{I}$, 
Definition \ref{QDS} requires both $\{{\cal T}^S_t\}$ and $\{{\cal
T}^F_t\}$ to be QDSs on their respective domain.  We begin with the 
following elementary Lemma:
\begin{lemma}\label{lemma} 
Let a linear operator $L:\Hi_1\otimes\Hi_2\rightarrow\Hi_3$ be
different from the zero operator.  Then there exist factorized pure
states in $\Hi_1\otimes\Hi_2\ominus\ker(L)$.\end{lemma} \proofn Assume
that $L\ket{\psi}= 0$ for all factorized
$\ket{\psi}\in\Hi_1\otimes\Hi_2$. Since such $\ket{\psi}$'s generate
the whole $\Hi_1\otimes\Hi_2,$ then by linearity it must be $L=0$ and
we conclude by contradiction.\hfill{\ \rule{0.5em}{0.5em}} %\qed

\vspace*{1mm}

%
%  MAIN INVARIANCE THEOREM
%

\begin{thm}[{Markovian invariance}] \label{markovianinvariance}
$\Hi_{S}$ supports an invariant subsystem
under Markovian evolution on $\Hi_I$ {iff}
for every initial state $\rho\in{\mathfrak
I}_{S}(\Hi_I),$ with $\rho_{S}\in{\mathfrak D}(\Hi_{S}),\,\rho_{F}\in{\mathfrak
D}(\Hi_{F}),$ the following conditions hold:
\begin{eqnarray}
\hspace*{-6mm}&&\ddt \rho (t) =\left(
\begin{array}{c|c}
 {\cal L}_{SF}(\rho_{SF}(t)) & 0  \\\hline
  0  &  0  
\end{array}
\right),\;\;\forall t\geq 0,\;\;\label{eq:LNF}\\ 
\hspace*{-6mm}&& {{\trace}}_F\left[{\cal L}_{SF} (\rho_{SF}(t))\right]={\cal L}_{S}(\rho_{S}(t)),\;\;\forall
t\geq0,\;\;\;
\label{partialF}
\end{eqnarray}
where ${\cal L}_{SF}$ and ${\cal L}_{S}$ are QDS generators on
$\Hi_S\otimes\Hi_F$ and $\Hi_S$, respectively.
\end{thm}

\proofn Since Definition \ref{invariance} is obeyed, computing the
infinitesimal generator of \eqref{eq:invariance} (at $t=0$) yields
\beq\label{eq:markoviangen}
\left. \ddt \rho (t) \right|_{t=0}
=\left(
\begin{array}{c|c}
( {\cal L}_{S}\otimes {\mathbb I}_F + {\mathbb I}_S \otimes {\cal
L}_F)(\rho_{S}\otimes\rho_F) & 0 \\
\hline 0 & 0
\end{array}
\right).\eeq 
\noindent
Then the time-invariant generator must have the form
\eqref{eq:LNF}. Take the partial trace over $\Hi_F$, and observe that
$\trace(\rho_F)=1$, $\trace({\cal L}_F(\rho_F))=0$. Then
\eqref{partialF} holds.

To prove the opposite implication, assume that \eqref{eq:LNF} and
\eqref{partialF} hold, and $\rho \in {\mathfrak I}_S(\Hi_I)$.  Since
$\rho$ evolves under a QDS generator that can be written in the form
\eqref{eq:lindblad}, with Hamiltonian $H$ and noise operators $L_k$
partitioned as in \eqref{eq:blocks}, computing the generator at a
generic time $t$ by blocks yields:
\begin{eqnarray*}
\ddt\rho=\left(
\begin{array}{c|c}
{\cal L}_{SF}(\rho) & {\cal L}_{P}(\rho)\\ \hline
{\cal L}_{Q}(\rho) & {\cal L}_{R}(\rho)\\
\end{array}
\right),
\end{eqnarray*}
where
\begin{eqnarray*}
{\cal L}_{SF}(\rho)&\hspace*{-2mm} = \hspace*{-2mm}&-i[H_{SF},\rho_{SF}]+\frac{1}{2}
\sum_{k}\left(2L_{SF,k}\rho_{SF}L_{SF,k}^\dag\right.\\
  &&-\left. 
  \left\{L_{SF,k}^\dag
L_{SF,k}+L_{Q,k}^\dag L_{Q,k},\rho_{SF}\right\}\right),\\ 
{\cal L}_{Q}(\rho) &\hspace*{-2mm} = \hspace*{-2mm}&-iH_P^\dag\rho_{SF}+\frac{1}{2}
\sum_{k}\left(2L_{SF,k}\rho_{SF} L_{Q,k}^\dag\right.\\
  &&- \left. 
  \rho_{SF}(L_{SF,k}^\dag
L_{P,k}+L_{Q,k}^\dag L_{R,k})\right)^\dag ,\\ 
{\cal L}_{P}(\rho)&\hspace*{-2mm} =\hspace*{-2mm}
 & i\rho_{SF}H_P+\frac{1}{2}\sum_{k}\left(2L_{SF,k}\rho_{SF}
L_{Q,k}^\dag\right.\\
  &&- \left. 
  \rho_{SF}(L_{SF,k}^\dag
L_{P,k}+L_{Q,k}^\dag L_{R,k}\right), \\ 
{\cal L}_{R}(\rho) & \hspace*{-2mm}= \hspace*{-2mm}
&\frac{1}{2}\sum_{k}2L_{Q,k}\rho_{SF}L_{Q,k}^\dag .
\end{eqnarray*}
%\label{eq:Lblocks}

\noindent 
In order to satisfy \eqref{eq:LNF}, it must be
$\frac{1}{2}\sum_{k}2L_{Q,k}\rho_{SF}L_{Q,k}^\dag=0$, for every
$\rho_{SF}=\rho_S\otimes\rho_F$. Consider for example pure product
states $\rho_{SF}=\ket{\psi}\bra{\psi}\otimes\ket{\phi}\bra{\phi}$.
Then, by observing that $L_{Q,k}\rho_{SF}L_{Q,k}^\dag$ is positive for
every $k$ and by using Lemma \ref{lemma}, it must be $L_{Q,k}=0$ for
every $k$.  Next, to ensure that $\rho_P(t)=0$ for every $t$, the
remaining contribution to ${\cal L}_{P}(\rho)$ must vanish, that is,
by using Lemma \ref{lemma} again, it must be: \beq\label{eq:offdiag}
iH_P-\frac{1}{2}\sum_{k}L_{SF,k}^\dag L_{P,k}=0.\eeq
\noindent 
This leaves a $SF$-block of the form:
\begin{eqnarray*}
-i[H_{SF},\rho_{SF}] &+&\frac{1}{2}\sum_{k}\left(2L_{SF,k}
\rho_{SF}L_{SF,k}^\dag\right. \\ 
& -& \left. 
\left\{L_{SF,k}^\dag
L_{SF,k},\, \rho_{SF}\right\}\right),
\end{eqnarray*}
\noindent 
which indeed satisfies \eqref{eq:LNF}. To see this, notice that we may
always write $L_{SF,k}=\sum_i M_{k,i}\otimes N_{k,i},$ with $M_{k,i}$
($N_{k,i}$) being operators on $\Hi_{S}$ ($\Hi_F$), respectively (this
follows e.g. from the operator
Schmidt-decomposition~\cite{nielsen-operatorschmidt}). Thus, we obtain
\beq
\begin{split} & \hspace*{-4mm} 2L_{SF,k}\rho_{SF}L_{SF,k}^\dag-\left\{L_{SF,k}^\dag
L_{SF,k},\rho_{SF}\right\}=\\=&
\sum_{i,j}\left(2M_{k,i}\rho_{S}M_{k,j}^\dag\otimes
N_{k,i}\rho_FN_{k,j}^\dag \right.\\ &\left. 
-(M_{k,j}^\dag
M_{k,i}\rho_{S}\otimes N_{k,j}^\dag 
N_{k,i}\rho_{F} \right.\\&\left.
+\rho_{S}M_{k,j}^\dag M_{k,i}\otimes
\rho_{F}N_{k,j}^\dag N_{k,i})\right).\end{split} \eeq
\noindent
By tracing over $\Hi_F$, and using the cyclic property of the (full)
trace, we obtain: 
\begin{eqnarray*}
\begin{split}
& \trace_F \left[ 2L_{SF,k}\rho_{SF}L_{SF,k}^\dag-\left\{L_{SF,k}^\dag
L_{SF,k},\rho_{SF}\right\} \right]= \label{eq:partial} \\&
\sum_{i,j}b_{ij}^k\left(2M_{k,i}\rho_{S}M_{k,j}^\dag
-(M_{k,j}^\dag M_{k,i}\rho_{S}+\rho_{S}M_{k,j}^\dag M_{k,i})\right),
\end{split}
\end{eqnarray*}
\noindent 
where $b_{ij}^k=\trace(\rho_F N_{k,j}^\dag N_{k,i})$.  Since we
require \eqref{partialF} to be {\em independent} of $\rho_F$,
$L_{SF,k}$ must take one of the following two forms:
\begin{eqnarray*}
&&L_{SF,k}=\sum_jM_{k,j}\otimes \mathbb{I}_F=L_{S,k}\otimes
\mathbb{I}_F,\\%\quad 
&&L_{SF,k}=\sum_{j}\mathbb{I}_{S}\otimes
N_{k,j}=\mathbb{I}_{S}\otimes L_{F,k}.
\end{eqnarray*}
\noindent
A similar reasoning shows that $H$ is constrained to
$$H=H_S\otimes \mathbb{I}_F +\mathbb{I}_S\otimes H_{F}.$$ 
\noindent
Thus, the generator has the form declared in \eqref{eq:markoviangen}.
\hfill{\ \rule{0.5em}{0.5em}} %\qed

%
% MATRIX BLOCK CONDITION FOR INVARIANCE
%

\vspace*{1mm}

As a byproduct, Theorem \ref{markovianinvariance}'s proof gives
explicit necessary and sufficient conditions for the blocks of $H$ and
$L_k$ to ensure invariance. We collect them in the following:

\begin{cor}[{Markovian invariance}]\label{invariancecondition} 
Assume that $\Hi_I=(\Hi_{S}\otimes\Hi_F)\oplus \Hi_R$, and let
$H,\,\{L_k\}$ be the Hamiltonian and the error generators of a
Markovian QDS as in \eqref{eq:lindblad}. Then $\Hi_{S}$ supports an
invariant subsystem iff $\forall\, k$:
\begin{eqnarray} 
&&L_{k}=\left(
\begin{array}{c|c}
  L_{S,k}\otimes L_{F,k} & L_{P,k} \nonumber \\\hline 0 & L_{R,k}
\end{array}
\right), \\ && iH_P-\frac{1}{2}\sum_k(L_{S,k}^\dag\otimes
L_{F,k}^\dag)L_{P,k}=0,\label{eq:nonrobust}\\ && H_{SF}=H_S\otimes
\mathbb{I}_F +\mathbb{I}_S\otimes H_{F},\nonumber 
\end{eqnarray} 
\noindent 
where for each $k$ either $L_{S,k}=\mathbb{I}_S$ or
$L_{F,k}=\mathbb{I}_F$ (or both). 
\end{cor}

\vspace*{1mm}

If we require parametric model robustness, as specified in Definition
\ref{robustproperty}, additional constraints emerge from imposing that
\eqref{eq:LNF} and \eqref{partialF} hold irrespective of parameter
uncertainties.  The results may be summarized as follows:

\vspace*{1mm}

\begin{thm}[{Robust Markovian invariance}]\label{GKSinvariance} 
Assume $\Hi_I=(\Hi_{S}\otimes\Hi_F)\oplus \Hi_R$. (i) Let $\{F_k\}$ be
the error generators in \eqref{eq:GKS}. Then $\Hi_{S}$ supports an
{\em $A$-robust invariant subsystem} iff $\forall j,k$,
\begin{eqnarray} && F_{k}=\left(
\begin{array}{c|c}
  F_{S,k}\otimes F_{F,k} & F_{P,k} \nonumber \\\hline 0 & F_{R,k}
\end{array}
\right), \\ 
&& F_{P,k}^\dag(F_{S,j}\otimes F_{F,j})=0,\label{eq:robust1}\\ 
&& H_{SF}=H_S\otimes \mathbb{I}_F
+\mathbb{I}_S\otimes H_{F},\quad H_P=0,\label{eq:HamBlocks}
\end{eqnarray}
where either $F_{S,k}=\mathbb{I}_S$ for every $k$, or
$F_{F,k}=\mathbb{I}_F$ for every $k$, or both.  (ii) If $\{L_k\}$ are
the noise operators in \eqref{eq:lindblad}, then $\Hi_{NS}$ supports a
{\em $\gamma$-robust invariant subsystem} iff $\forall \, k$,
\begin{eqnarray} && L_{k}=\left(
\begin{array}{c|c}
 L_{S,k}\otimes L_{F,k} & L_{P,k}    \nonumber  \\\hline
 0  &  L_{R,k}   
\end{array}
\right),\\
&& L_{P,k}^\dag(L_{S,k}\otimes L_{F,k})=0,\\
&& H_{SF}=H_S\otimes \mathbb{I}_F
+\mathbb{I}_S\otimes H_{F},\; H_P=0,
\end{eqnarray}
and for each $k$, either $L_{S,k}=\mathbb{I}_S$ or
$L_{F,k}=\mathbb{I}_F$ (or both).\end{thm}

\vspace*{1mm}

\proofn Consider case (i). Given Theorem \ref{markovianinvariance},
conditions \eqref{eq:LNF}-\eqref{partialF} must hold irrespective
of $A=\left(a_{jk}\right)$ in \eqref{eq:GKS}. The lower-diagonal block
is now $\sum_{jk}a_{jk} F_{Q,j}\rho_{SF}F_{Q,k}^\dag.$ Considering
only the diagonal terms $j=k$, we are again led to require $F_{Q,k}=0$
for every $k$.  Thus, condition \eqref{eq:offdiag} must be replaced by
$iH_P-\sum_{jk}a_{jk}F_{SF,k}^\dag F_{P,j}=0,$
\noindent 
which is true for every $A=\left(a_{jk}\right)$ iff $H_P$ and
$F_{P,k}^\dag(F_{S,j}\otimes F_{F,j})$ vanish independently, as stated
in \eqref{eq:robust1}, \eqref{eq:HamBlocks}.

To complete the proof, as before we write $F_{SF,k}=\sum_{k,l}
M_{k,l}\otimes N_{k,l},$ with $M_{k,l}$ ($N_{k,l}$) operators on
$\Hi_{NS}$ ($\Hi_F$), respectively. Then we have: 

\begin{eqnarray*}
\begin{split}
&2F_{SF,j}\rho_{SF}F_{SF,k}^\dag-\left\{F_{SF,k}^\dag
F_{SF,j},\rho_{SF}\right\}=\\
&=\sum_{l,m}\left(2M_{j,l}\rho_{NS}M_{k,m}^\dag\otimes
N_{j,l}\rho_FN_{k,m}^\dag \right.\\ 
&\hspace*{4mm} - \left.(M_{k,m}^\dag
M_{j,l}\rho_{NS}\otimes N_{k,m}^\dag N_{j,l}\rho_{F}\right.\\
&\hspace*{4mm} +\left. \rho_{NS}M_{k,m}^\dag M_{j,l}\otimes \rho_{F}N_{k,m}^\dag
N_{j,l})\right) .\end{split} 
\end{eqnarray*}
\noindent
By tracing over $\Hi_F$ and using cyclicity yields
\begin{eqnarray*}
\label{eq:partial1}
\sum_{l,m}\hspace*{-1mm}b_{lm}^{kj}\hspace*{-1.5mm}\left(\hspace*{-0.5mm}2M_{j,l}\rho_{NS}M_{k,m}^\dag
\hspace*{-1.5mm}-(M_{k,m}^\dag M_{j,l}\rho_{NS}+\rho_{NS}M_{k,m}^\dag
M_{j,l})\hspace*{-1mm}\right),
\end{eqnarray*}
\noindent 
where $b_{lm}^{kj}=\trace(\rho_F N_{k,m}^\dag N_{j,l})$.  Since we wish
\eqref{partialF} to be independent of $\rho_F$, it follows that for
every $k$, $F_{SF,k}$ takes one of two possible forms:
$$F_{SF,k}=\sum_m\mathbb{I}_{NS}\otimes N_{m,k}=\mathbb{I}_{NS}\otimes
F_{F,k},%\quad 
$$
$$ F_{SF,k}=\sum_{l}M_{k,l}\otimes \mathbb{I}_F=F_{NS,k}\otimes
\mathbb{I}_F,$$ 
\noindent 
which establishes the conditions for $A$-robustness.

For $\gamma$-robustness, it suffices to specialize the above proof to
diagonal $A$, that is, to consider only $j=k$. This let us identify
the $L_{k}$'s with the $F_{k}$'s, and the result follows.  \hfill{\
\rule{0.5em}{0.5em}} %\qed

%
%  NOISELESS SECTION 
%

\vspace*{-2mm} 

\subsection{Noiseless subsystems}

As remarked after Definition \ref{initialization}, given
$\rho\in{\mathfrak D}(\Hi_I)$, the reduced projected state:

\vspace*{-3mm}

\beq\label{eq:rhobar}\bar\rho_{S}=\trace_{F}(\Pi_{SF}\rho
\Pi_{SF}^\dag),\eeq
\noindent
need not be a valid reduced state of ${\cal S}$ if $\rho_R\neq 0$,
since its trace might be less than one.  Still, to the purposes of
defining noiseless behavior, there is no reason for requiring that the
evolution has to be confined to ${\mathfrak I}_S({\cal H}_{I})$, as
long as the information encoded in the intended subsystem is
preserved, i.e. it undergoes unitary evolution. This motivates a {\em weaker} definition of initialization:

\vspace*{1mm}
\begin{defin}[{%Perfect 
Reduced state initialization}]\label{pinitialization} 
The system $\mathcal{I}$ with state $\rho$ is {\em %perfectly
initialized in $\Hi_{SF}$ with reduced state $\bar\rho_{S}\in{\mathfrak
D}(\Hi_{S})$} if the blocks of $\rho$ satisfy:
\begin{itemize}   
\item[{(i$'$)}] %$\rho_{SF}\in{\frak D}(\Hi_{SF})$, such that
$\trace_F(\rho_{SF})=\bar\rho_{S}$;
\item[{(ii)}] $\rho_P=0,\rho_R=0.$
\end{itemize}
\end{defin}

\vspace*{1mm}

{\em Remark:} The above definition is equivalent to state
initialization as in given in Definition \ref{initialization} if we
restrict to {\em pure} states in $\Hi_S$, but it allows for entangled
states otherwise. 

%A general definition of NS for a dynamical map may then be phrased as
%follows:

\begin{defin}[{Noiselessness}] \label{NSproperty} Let ${\cal I}$ 
evolve under TPCP maps.  A subsystem $\cal{S}$ is a NS for the
evolution if for every initial state $\rho(0)$ of $\cal{I}$ 
initialized in $\cal{S}$ with reduced state $\bar\rho_{S}(0)$:
\beq
\bar\rho_{NS}(t)=U(t)\bar\rho_{S}(0)U(t)^\dag , \;\;\; t \geq 0, \eeq
\noindent 
where $U(t)$ is a unitary operator on $\Hi_{S}$, {\em independent of
the initial state} on $\Hi_{I}$. If ${\cal H}_F \simeq {\mathbb C}$,
the NS ${\cal S}$ reduces to a DFS.
\end{defin}

Following ~\cite{lidar-initializationDFS}, we shall say
that {\em perfect NS initialization} occurs when (i$'$) and (ii) are
obeyed, and call subsystem ${\cal S}$ {\em imperfectly initialized}
whenever (i$'$) or (ii) is violated.\ Physically, Definition 8
requires the state component $\bar \rho_{NS}$ carried by ${\cal H}_S$
to {\em evolve unitarily}, independently of the rest.  The following
proposition establishes how, in fact, a perfecly initialized NS is a
special case of an invariant subsystem:
% if Definition 7 holds:
%perfect reduced state initialization holds.

\begin{prop}\label{NSlink} Let $\Hi_{I}=(\Hi_{NS}\otimes \Hi_{F})\oplus\Hi_R.$ Then $\Hi_{NS}$
supports a NS for some given TPCP dynamics {iff} for all initial condition $\rho\in{\mathfrak I}_{NS}(\Hi_I)$, with $\rho_{NS}\in{\mathfrak
D}(\Hi_{NS}),\,\rho_{F}\in{\mathfrak D}(\Hi_{F})$, the evolved state of
${\cal I}$ obeys
\begin{equation}\label{eq:NSproperty}
\rho(t)=\left(
\begin{array}{c|c}
  U(t)\rho_{NS}  U^\dag(t)\otimes{\cal T}^F_t(\rho_F  ) & 0    \\\hline
  0  &  0  
\end{array}
\right),\;\; t \geq 0, 
\end{equation}
where $U(t)$ is a unitary operator on $\Hi_{NS}$ and $\{{\cal T}^F_t\}$ are TPCP maps on $\Hi_F$ alone.
\end{prop}

\proofn %Since for NS reduced state initialization is 
Assume $\Hi_{NS}$ to support an NS and to be initialized with reduced state $\bar\rho_{NS},$ where it could be
$\rho_{NSF}(0)\neq \bar\rho_{NS}(0)\otimes\rho_F(0).$ Let
$\rho_{NSF}(0)=\sum_k S_k\otimes F_k$, with $S_k$ and $F_k$ operators
on $\Hi_S$ and $\Hi_F,$ respectively.  Thus, $\bar\rho_{NS}(0)=\sum_k
\trace(F_k)S_k.$ Notice that, by linearity: \beqan
&&\hspace*{-7mm}\trace_F(\Pi_{SF}\rho(t)\Pi_{SF}^\dag)=%\hspace*{-3mm}=\hspace*{-2mm}
\trace_F\Big[\sum_k U(t)S_k U^\dag (t)\otimes {\cal T}^F_t(F_k)\Big]\\&&
\hspace*{-5mm}=
%\hspace*{-3mm}=\hspace*{-2mm}
\sum_k U(t)S_k U^\dag(t)\, \trace\Big({\cal T}^F_t(F_k)\Big)
%&\hspace*{-3mm}=\hspace*{-2mm}&
%U(t)\Big(\sum_k\trace(F_k)S_k\Big) U^\dag (t)\\ 
=%\hspace*{-3mm}=\hspace*{-2mm}
U(t)\bar\rho_{NS}(0) U^\dag (t). \eeqan
\noindent 
Thus, the condition is sufficient. 

To prove the other implication, notice that if the evolution of
$\bar\rho_{NS}$ is unitary, it preserves the trace. By the properties
of partial trace and by \eqref{eq:rhobar}, it then follows that
$\trace (\bar\rho_{NS}(t))=\trace(\rho_{NSF}(t)).$ Therefore,
$\rho_R=0$ must vanish at all times in order to ensure TP-evolution in
the $SF$-block and, similarly, $\rho_P=0$ in order to guarantee
positivity of the whole state \footnote{This follows from the fact
that if $\rho=(\rho_{ij})\geq 0,$ then $|\rho_{ij}|\leq
\sqrt{\rho_{ii}\rho_{jj}}$.}. Then for every $t \geq 0$ the evolution
must take the form:
$$
\rho(t)=\left(
\begin{array}{c|c}
 {\cal T}^{NSF}_t(\rho_{NSF} ) & 0    \\\hline
  0  &  0  
\end{array}
\right),$$ 
\noindent 
where ${\cal T}^{NSF}_t$ is a TPCP map on $\Hi_{NS}\otimes\Hi_F$. Now
use the Kraus representation theorem and the operator-Schmidt
decomposition, by employing a basis for ${\mathfrak B}(\Hi_{S}),$ say
$\{M_j\}$, such that, $\forall\,\rho_{NS}\in{\mathfrak
D}(\Hi_{NS}),\,\rho_{F}\in{\mathfrak D}(\Hi_{F})$,
$${\cal T}^{NSF}_t(\rho_{NS}\otimes\rho_{F})=\sum_{klm}M_l \rho_{NS}
M_m^\dag\otimes N_{k,l}\rho_F N_{k,m}^\dag .$$ 

\vspace*{-2mm}

\noindent 
Thus,
$$\trace_F[\Pi_{NSF}\rho(t)\Pi_{NSF}^\dag]={\cal T}^{S}_t(\rho_{NS})
=\sum_{lm}\alpha_{lm} M_l\rho_{NS} M_m^\dag ,$$

\vspace*{-2mm}

\noindent
where $\alpha_{lm}=\trace\left( \sum_{k} N_{k,m}^\dag
N_{k,l}\rho_F\right)$ is a positive matrix.  By exploiting the fact
that $\{M_j\}$ is a basis, and decomposing $\sum_{k} N_{k,m}^\dag
N_{k,l}$ in Hermitian and skew-Hermitian parts, one can see that a
necessary and sufficient condition in order for ${\cal T}^{S}_t$ to be
independent of $\rho_F$ is that $\sum_{k}N_{k,m}^\dag N_{k,l}=
\alpha_{ml}\mathbb{I}_F\,$ for every $j,k.$ By imposing that
$\trace_F[\Pi_{SF}\rho(t)\Pi_{SF}^\dag]=
U(t)\bar\rho_{S}(0)U^\dag(t)$, $(\alpha_{jk})$ must have rank one,
thus $\alpha_{jk}=\alpha_j\alpha_k$ for some $\{\alpha_j\}$. If we
additionally choose the operator basis $\{M_j\}$ so that $M_1=U(t)$,
then $\alpha_{11}$ is the only non-zero entry, in particular,
$\sum_{k}N_{k,l}^\dag N_{k,l}=0$ for every $l\neq 1$.  This implies
$N_{k,l}=0$ for every $l\neq 1$, thus yielding to the desired conclusion:
$${\cal T}^{SF}_t(\rho_{S}\otimes\rho_{F})=U(t)\rho_S U(t)^\dag\otimes
\sum_{k} N_{k}\rho_F N_{k}^\dag .$$ 

\vspace*{-4mm}

\hfill{\ \rule{0.5em}{0.5em}}

\vspace*{1mm}

On one hand, as a consequence of the above Proposition, if reduced
state initialization is assumed (as in Definition
\ref{pinitialization}), the factor $\Hi_{NS}$ {\em supports an NS only
if it is invariant}. On the other hand, under the stronger condition
of initialization of Definition \ref{initialization}, if $\Hi_{NS}$ is
{\em invariant and unitarily evolving}, then it supports a NS.
Accordingly, most of the results concerning NSs may be derived as a
specialization of conditions for invariance. Remarkably, this also
implies that in the particular case of a NS, the invariance property
is robust with respect to the initialization in the $NSF$-block, that
is, condition (i) may be effectively relaxed to (i$'$).  This is {\em
not} true for general invariant subsystems.  Explicit
characterizations of the Markovian noiseless property may then be
established as summarized in the rest of this Section.

\vspace*{1mm}

\begin{cor}[{Markovian NS}] \label{NS}% Let ${\cal I}$ 
%with state $\rho$ be initialized in $\Hi_{NS}$ with reduced state
%$\bar{\rho}_{NS}$. 
Let $\Hi_{I}=(\Hi_{NS}\otimes \Hi_{F})\oplus\Hi_R.$ 
Then $\Hi_{NS}$ supports a NS under Markovian
evolution on ${\cal H}_I$ iff for every initial state $\rho\in{\mathfrak I}_{NS}(\Hi_I),$ with
$\rho_{NS}\in{\mathfrak
D}(\Hi_{NS}),\,\rho_{F}\in{\mathfrak D}(\Hi_{F}),$ and $\forall t\geq 0$:
\begin{eqnarray}
\hspace*{-10mm}
&&\ddt \rho (t) =\left(
\begin{array}{c|c}
 {\cal L}_{NSF}(\rho_{NSF}(t)) & 0  \\\hline
  0  &  0  
\end{array}
\right),\;\;\;\label{eq:LNF1}\\ 
\hspace*{-10mm}
&& {\trace}_F\left[{\cal L}_{NSF}(\rho_{NSF}(t) )
\right]=-i[H_{NS},\rho_{NS}(t)],\;\;\;\label{partialF1}
\end{eqnarray}
where ${\cal L}_{NSF}$ and ${\cal L}_{NS}$ are QDS generators on
$\Hi_{NS}\otimes\Hi_F$ and $\Hi_{NS}$, respectively.
\end{cor}

\proofn Given Proposition \ref{NSlink} and Theorem
\ref{markovianinvariance}, we need to ensure that the evolution in
$\Hi_{NS}$ is unitary. That is, the $NSF$-block must be driven by an
generator of the form $-i[H_{NS},\rho_{NS}]\otimes \rho_F
+\rho_{NS}\otimes {\cal L}_F(\rho_F)$, which replaces
\eqref{eq:markoviangen} and ensures unitary evolution on the
NS-factor, while allowing for general non-unitary Markovian dynamics
on $\Hi_F$. The proof of Theorem \ref{markovianinvariance} applies,
with the noise operators constrained to have the form
$$\hspace{15mm}L_{NSF,k}=\sum_{j}\mathbb{I}_{S}\otimes
N_{k,j}=\mathbb{I}_{S}\otimes L_{F,k}. $$

\vspace*{-4mm}

\mbox{}\hfill{\ \rule{0.5em}{0.5em}}
%$$ %\qed

Accordingly, the necessary and sufficient conditions on the matrix
blocks of $H$ and $L_k$ for NS-behavior are modified to:

\vspace*{1mm}

\begin{cor} \label{NScondition} 
Assume $\Hi_I=(\Hi_{NS}\otimes\Hi_F)\oplus \Hi_R$, and let $H$,
$\{L_k\}$ be the Hamiltonian and the error generators of a Markovian
QDS as in \eqref{eq:lindblad}. Then $\Hi_{NS}$ supports a NS iff
$\forall k$:
\begin{eqnarray} &&L_{k}=\left(
\begin{array}{c|c}
  \mathbb{I}_{NS}\otimes L_{F,k} & L_{P,k} \nonumber \\
\hline 0 & L_{R,k}
\end{array}
\right),\\ &&
iH_P-\frac{1}{2}\sum_k(\mathbb{I}_{NS}\otimes L_{F,k}^\dag)L_{P,k}=0,\\ 
&& H_{NSF}=H_{NS}\otimes \mathbb{I}_F +\mathbb{I}_{NS}\otimes H_{F}.
\nonumber 
\end{eqnarray} \end{cor}

\vspace*{1mm}

While derivations differ, Corollary \ref{NScondition} provides the
same NS-characterization of Theorem 5 in
\cite{lidar-initializationDFS}.  Beside more directly tying to the CP
context, our approach shows how the NS notion may emerge as a
specialization of the conditions for invariance.  Following the same
lines as in Section II-B, we next proceed to a general result for{\em
$A$- and $\gamma$-robust NSs}, which completes the partial conditions
proposed in \cite{kempe-NS}:

\begin{cor}[{Robust Markovian NS}]\label{GKSNS} Assume that
$\Hi_I=(\Hi_{NS}\otimes\Hi_F) \oplus \Hi_R$. (i) Let $\{F_k\}$ be the
error generators in \eqref{eq:GKS}. Then $\Hi_{NS}$ supports an 
{$A$-robust NS} iff $\forall j,k$:
\begin{eqnarray} && F_{k}=\left(
\begin{array}{c|c}
  \mathbb{I}_{NS}\otimes F_{F,k} & F_{P,k} \nonumber \\\hline 0 & F_{R,k}
\end{array}
\right),\\ && F_{P,k}^\dag(\mathbb{I}_{NS}\otimes F_{F,j})=0, \\ &&
H_{NSF}=H_{NS}\otimes \mathbb{I}_F +\mathbb{I}_{NS}\otimes H_{F},\quad
H_P=0.\label{eq:HamBlocks1}
\end{eqnarray}
(ii) If $\{L_k\}$ are the error generators in \eqref{eq:lindblad},
then $\Hi_{NS}$ supports a $\gamma$-robust NS iff $\forall k$:
\begin{eqnarray} && L_{k}=\left(
\begin{array}{c|c}
  \mathbb{I}_{NS}\otimes L_{F,k} & L_{P,k}     
  \\\hline
  0  &  L_{R,k}   
\end{array}
\right),\quad \forall k,\\
&& L_{P,k}^\dag(\mathbb{I}_{NS}\otimes L_{F,k})=0, \\ &&
H_{NSF}=H_{NS}\otimes \mathbb{I}_F +\mathbb{I}_{NS}\otimes H_{F},\;
H_P=0.
\end{eqnarray}
\end{cor}

\vspace{1mm}

\proofn Given Theorem \ref{GKSinvariance}, it suffices to ensure that
evolution in $\Hi_{NS}$ be unitary (as in Corollary \ref{NS}), for
every $A$. This is true {iff} $H_P=0$ and
$F_{P,k}^\dag(\mathbb{I}_{NS}\otimes F_{F,j})=0$ independently.  From
\eqref{eq:partial1}, $F_{SF,k}=\mathbb{I}_{S}\otimes F_{F,k}$ must
hold for every $k$.  The specialization to a $\gamma$-robust NS
follows from similar observations. \hfill{\ \rule{0.5em}{0.5em}}

\vspace*{1mm}

Clearly, an $A$-robust NS may exist only if the $\{F_k\}$ do not
generate the whole ${\mathfrak B}(\Hi_I)$, that is, we are
restricting to a set of possible noise generators as remarked in
Section \ref{sec:robustness}.
For applications, it may be useful to further specialize the result to
the case of a $\gamma$-robust DFS, for which $\Hi_F$ is trivial:

\begin{cor}[{$\gamma$-robust DFS}] \label{lindbladDFS} 
Assume $\Hi_I=\Hi_{DFS} \oplus \Hi_R$. Let $\{L_k\}$ be the error
generators in \eqref{eq:lindblad}. Then $\Hi_{DFS}$ is a
{$\gamma$-robust} DFS iff $\forall k$: 
%$$ H_P=0,$$
\beq
 H_P=0,\quad L_{k}=\left(
\begin{array}{c|c}
  c_k\mathbb{I}_{DFS} & L_{P,k}     \\\hline
  0  &  L_{R,k}   
\end{array}
\right),\label{eq:ii} \eeq 
\noindent
with $L_{P,k}=0$ if $c_k\neq0$.
\end{cor} \proofn In Corollary \ref{GKSNS} above, set
$\Hi_{F}=\C$. \hfill{\ \rule{0.5em}{0.5em}} %\qed

\vspace*{1mm}

An alternative formulation of Corollary \ref{lindbladDFS} also holds:

\begin{prop}[{Alternative $\gamma$-robust DFS condition}] 
$\Hi_{DFS}=\mbox{span}\{\ket{\phi_j^{DFS}}\}$ is a {$\gamma$-robust
Markovian DFS} subspace of $\Hi_I$ iff $\forall j,k$ the following
conditions hold:
$$H_P=0, \quad L_{k}\ket{\phi_j^{DFS}}=c_k\ket{\phi_j^{DFS}},$$ 
$$L_{k}^\dag
L_{k}\ket{\phi_j^{DFS}}=|c_k|^2\ket{\phi_j^{DFS}}.\, $$ 
\end{prop}
\vspace*{1mm}
\proofn Assume that $\forall k$,
$L_{k}\ket{\phi_j^{DFS}}=c_k\ket{\phi_j^{DFS}}.$ %\noindent
Then it must be 

$$L_{k}=\left(
\begin{array}{c|c}
  c_k\mathbb{I}_{DFS} & L_{P,k}     \\\hline
  0  &  L_{R,k}   
\end{array}
\right). $$ 
\noindent
Since 
$$L_{k}^\dag L_{k}=\left(
\begin{array}{c|c}
  c_kc_k^*\mathbb{I}_{DFS} & c_k^*L_{P,k}     \\\hline
  c_kL_{P,k}^\dag  & L_{P,k}^\dag L_{P,k}+L_{R,k}^\dag L_{R,k}   
\end{array}
\right),$$ 
\noindent
then $ L_{k}^\dag L_{k}\ket{\phi_j^{DFS}}=|c_k|^2\ket{\phi_j^{DFS}}$
is true {iff} the conditions of the proof of Corollary
\ref{lindbladDFS} are obeyed.  \hfill{\ \rule{0.5em}{0.5em}} %\qed

\vspace*{1mm}

{\em Remark:} According to the above Corollary, $\gamma$-robust
DFS-states are joint (right) eigenvectors of {\em both} each Lindblad
operator $L_{k}$ and each ``jump'' operator $L_{k}^\dag L_{k}$.  Such
characterizations of $A$- and $\gamma$-robust DFSs link our analysis
to the definition presented in \cite{whaley-DFS}. In fact, the
definition of DFS property invoked there imposes more constraints than
our Definition \ref{pinitialization}: it requires the Hamiltonian to
preserve the DFS {\em independently} from the dissipative component of
the generator.  This may be regarded as a yet different kind of
robustness, weaker than both $\gamma$- and $A$-robustness investigated
here.

\iffalse\footnote{However, in \cite{whaley-DFS} the argument that the
NS condition in \cite{lidar-initializationDFS}, which equivalent to
ours, is not sufficient to ensure the DFS behavior is based on the
Counterexample 21. In that model, the noisy evolution is crucially
dependent on a sign error in Equation (37), in contrast with what has
been claimed in the footnote [18].}.  \fi

\vspace*{-2mm} 

\subsection{Imperfect initialization}\label{imperfect}

Thus far, we have addressed model robustness of the invariance and the
noiselessness properties.  As the relevant subsystem dynamics also
depends on the initial state, it is natural to ask how critical
initialization is to the purposes of ensuring a desired behavior.
This motivates the introduction of a different robustness notion:

\begin{defin} \label{staterobust} 
Assume that $\Hi_I = (\Hi_S \otimes \Hi_F)\oplus \Hi_R$, and let
${\cal I}$ undergo QDS dynamics with a generator of the form
\eqref{eq:GKS} or \eqref{eq:lindblad}.  Let a given property
${\mathfrak P}_S$ hold for the dynamical model %perfectly 
initialized
in $\Hi_{S}$, according to Definition \ref{initialization}.  If
${\mathfrak P}_S$ holds for
$\bar\rho_S=\textrm{{trace}}_F(\Pi_{SF}\rho\Pi_{SF})$ for every $\rho
\in{\frak D}(\Hi_I)$, then ${\mathfrak P}_S$ is said to be {\em
{$\rho$-robust}}.
\end{defin}

%\vspace*{1mm}

This approach leads to the same conditions for
imperfect-initialization Markovian NSs ({\em initialization-free NS})
obtained in \cite{lidar-initializationDFS}.  By Definition
\ref{staterobust}, considering $\Hi_I = (\Hi_{NS} \otimes \Hi_F)\oplus \Hi_R,$ a subsystem  supported on $\Hi_{NS}$  is a $\rho$-robust NS if for every $t\geq0$, $\forall\,\rho(0)$ with reduced state $\bar\rho_{NS}(0)\in{\mathfrak
D}(\Hi_{NS})$:
$$\bar\rho_{NS}(t)=
\trace_F(\Pi_{NSF}\rho(t)\Pi_{NSF}^\dag)=U(t)\bar\rho_{NS}(0)U^\dag(t),
$$
\noindent
where $U(t)$ is unitary on $\Hi_{NS}$. This means that the dynamics of
the $NSF$-block cannot be influenced by $\rho_P,\rho_R$.  Notice that
Proposition \ref{NSlink} has already clarified how imperfect
initialization in the $NSF$ block does not affect the unitary
character of the evolution of the reduced state of the NS. Computing the generator ${\cal L}(\rho)$ by blocks, by inspection
one sees that for all $k$ it must be $L_{P,k}=0$, $L_{Q,k}=0$, hence
$H_P=0$.  By the proof of Theorem \ref{GKSNS}, it also follows that
$L_{NSF,k}=\mathbb{I}\otimes M_k$.  Thus, the main difference with
respect to the perfect NS-initialization case is the {constraint
$L_{P,k}=0$}, which decouples the evolution of the $NSF$-block from
the rest. Notice that this also automatically ensures
$\gamma$-robustness in our framework.

\vspace*{-2mm} 

\subsection{Attractive subsystems}

The analysis developed so far indicates how initialization
requirements may be relaxed by requiring $\rho$-robustness.  However,
this implies in general tighter conditions on the noise operators,
which may be demanding to ensure and leave less room for Hamiltonian
compensation of the noise action (see Section \ref{control}).  In
order to both address situations where such extra constraints are not
met, as well as a question which is interesting on its own, we explore
conditions for a NS to be not only invariant, but also attractive:

\begin{defin}[{Attractive Subsystem}] \label{attraction} 
Assume that $\Hi_I=(\Hi_{S}\otimes\Hi_F) \oplus \Hi_R$. Then $\Hi_{S}$
supports an attractive subsystem with respect to a family $\{{\cal
T}_t\}_{t\geq 0}$ of TPCP maps if $\forall\rho\in{\mathfrak
D}(\Hi_{I})$ the following condition is asymptotically obeyed:
\begin{equation}\label{eq:attraction}
\lim_{t\rightarrow \infty}\left({\cal T}_t(\rho)-\left(
\begin{array}{c|c}
  \bar\rho_{S}(t)\otimes\bar\rho_F(t) & 0    \\\hline
  0  &  0  
\end{array}
\right)\right)=0,\end{equation} 
\noindent
where $\bar\rho_{S}(t)=\trace_{F}[\Pi_{SF}{\cal T}_t(\rho)
\Pi_{SF}^\dag],$ $\bar\rho_{F}(t)=\trace_{S}[\Pi_{SF}{\cal T}_t(\rho)
\Pi_{SF}^\dag].$
\end{defin}

An attractive subsystem may be thought of as a subsystem that
``self-initializes'' in the long-time limit, by somehow reabsorbing
initialization errors. Although such a desirable behavior only emerges
asymptotically, for QDSs one can see that convergence is exponential,
as long as some eigenvalues of ${\cal L}$ have strictly negative
real part.

We begin with a negative result which, in particular, shows how the
initialization-free and attractive characterizations are mutually
exclusive.

\vspace*{1mm}

\begin{prop}\label{hermitianL} Assume 
$\Hi_I=(\Hi_{NS}\otimes\Hi_F)\oplus \Hi_R,$ $\Hi_R\neq 0$, and let
$H,\,\{L_k\}$ be the Hamiltonian and the error generators as in
\eqref{eq:lindblad}, respectively.  Let $\Hi_{NS}$ support a NS. If
$L_{P,k}=L_{Q,k}^\dag=0$ for every $k$, then $\Hi_{NS}$ is not
attractive.\end{prop} \proofn Consider a block-diagonal state of the
form:
$$
\rho_B=\left(
\begin{array}{c|c}
  \rho_{SF} & 0    \\\hline
  0  &  \rho_R 
\end{array}
\right).$$
\noindent
It is straightforward to see that the generator 
has the form
$$\ddt\rho_B= \left(
\begin{array}{c|c}
  {\cal L}_{SF}(\rho_{SF}) & 0 \\\hline 0 & {\cal L}_{R}(\rho_R)
\end{array}
\right),$$ 
\noindent
which preserves the trace of $\rho_R$. Thus, if $\rho_R\neq 0$,
$\rho_B$ does not satisfy Eq.~\eqref{eq:attraction}.  \hfill{\
\rule{0.5em}{0.5em}} %\qed

\vspace*{1mm}

{\em Remark:} The conditions of the above Proposition are obeyed, in
particular, for NSs in the presence of purely {\em Hermitian} noise
operators, that is, $L_k=L_k^\dag$, $\forall k $.  As a consequence,
attractivity is never possible for this kind of {\em unital} Markovian noise, as
defined by the requirement of preserving the fully mixed state.
Still, even if the condition $L_{P,k}=L_{Q,k}^\dag=0$ condition holds,
attractive subsystems may exist in the pure-factor case, where
$\Hi_R=0$. Sufficient conditions are provided by the following:
\iffalse
The result is indeed negative, but the fact we have to require
$\Hi_R\neq 0$ still allows for attractive, purely-factor subsystems
(i.e. $\Hi_R=0$) even if the $L_p=L_q^\dag=0$ condition holds. In
fact, we provide here a result that gives sufficient conditions for
attraction in the $\Hi_I=\Hi_S\otimes\Hi_F$ case.
\fi

\begin{prop}\label{attractionfactor} Assume 
$\Hi_I=\Hi_{S}\otimes\Hi_F$ ($\Hi_R= 0$), and let $\Hi_{S}$ be
invariant under a QDS of the form
$${\cal L}= {\cal L}_{S}\otimes {\mathbb I}_F + {\mathbb I}_S \otimes{\cal
  L}_F .$$ 
\noindent
If ${\cal L}_F(\cdot)$ has a unique attractive state $\hat\rho_F$, then
$\Hi_S$ is attractive. \end{prop}

\proofn Let $\rho$ be a generic state on $\Hi_I=\Hi_{S}\otimes\Hi_F$.
$\rho$ may always be expressed (recall the proof of Theorem
\ref{markovianinvariance}) in the form $\rho=\sum_i P_i\otimes Q_i$.
Without loss of generality, we may take the $Q_i$ to be Hermitian. If
this is not the case, decompose $Q_i=Q_i^H+iQ_i^A$ into Hermitian and
skew-Hermitian parts, so that $P_i\otimes Q_i=P_i\otimes
Q^H_i+(iP_i)\otimes Q^A_i$.  Each of the $Q_i^{H,A}$ may be further
decomposed in a positive and a negative part, which one may normalize
to unit trace. To do so, consider the spectral representation of each
$Q_i$, separate the positive and negative eigenvalues, and partition
the matrix in a sum of two $ Q_i= Q^+_i+Q^-_i$. Normalize
$Q^+_i,\,Q^-_i$ to trace $1,\,-1,$ respectively, and reabsorb the
normalization coefficients and the minus sign, in $P_i$.  Thus, we can
write $\rho=\sum_i \tilde{P}_i\otimes \rho_{F,i},$ and
$\bar\rho_{NS}=\sum_i \tilde{P}_i$. By applying the above generator to
such a state and using the linearity of the evolution,
\[\begin{split}\lim_{t\rightarrow\infty}\rho_t & =\sum_i\lim_{t\rightarrow\infty}
 \left({\cal T}^t_{S}(\tilde{P}_i)\otimes {\cal T}^t_{F}(\rho_{F,i}) \right) \\&
 =\Big(\sum_i  \lim_{t\rightarrow\infty}{\cal
T}^t_{S}(\tilde{P}_i)\Big)\otimes \hat\rho_{F} 
=\lim_{t\rightarrow\infty}{\cal T}^t_{S}(\bar\rho_{NS})\otimes
\hat\rho_{F},
\end{split}\] 
thus the desired conclusion follows.\hfill{\ \rule{0.5em}{0.5em}}%\qed

%\vspace*{1mm}

In the mathematical-physics literature, QDSs with a unique attractive
stationary state are called {\em relaxing}, and have been mostly
studied in the '70 in the context of rigorous approaches to quantum
thermodynamics. Useful linear-algebraic conditions for determining
whether a generator ${\cal L}_F(\cdot)$ is relaxing are presented in
\cite{spohn-approach,spohn-equilibrium}.  The uniqueness of the
stationary state turns out to be {\em necessary} when considering NSs:

\begin{prop} Assume $\Hi_I=\Hi_{S}\otimes\Hi_F$, ($\Hi_R= 0$), 
%$\textrm{{dim}}(\Hi_S)>2$, 
and let $\Hi_{S}$ support a NS under a QDS
of the form
$${\cal L}=
 {\cal L}_{S}\otimes {\mathbb I}_F +{\mathbb I}_{S}\otimes{\cal L}_F.$$ 
\noindent 
If ${\cal L}_F(\cdot)$ admits at least two invariant states, then
$\Hi_S$ is not attractive. \end{prop}

\proofn It suffices to construct a state of the form:
$$\rho=p\rho_{S}^{(1)}\otimes\rho_{F}^{(1)}+(1-p)\rho_{S}^{(2)}\otimes\rho_{F}^{(2)},$$
\noindent
where $\rho_{S}^{(1)},\rho_{S}^{(2)}$ are orthogonal pure states on $\Hi_S$, and
$\rho_{F}^{(1)},\rho_{F}^{(2)}$ are the two invariant states for
${\cal L}_F$, and $0<p<1$. Again, by using the linearity of the
evolution, 
\begin{eqnarray*}
\rho(t)&\hspace*{-3mm}=\hspace*{-3mm}&[p{\cal T}_{S}(\rho_{S}^{(1)})
\otimes {\cal T}_{F}(\rho_{F}^{(1)})
\hspace*{-0.5mm} +\hspace*{-0.5mm} (1-p){\cal
T}_{S}(\rho_{S}^{(2)})\otimes {\cal T}_{F}(\rho_{F}^{(2)})]\\
&\hspace*{-3mm}
=\hspace*{-3mm}&pU_{S}(t)\rho_{S}^{(1)}U_S^\dag(t)\otimes
\rho_{F}^{(1)}
\hspace*{-0.5mm} +\hspace*{-0.5mm}
(1-p)U_{S}(t)\rho_{S}^{(2)}U_S^\dag(t)\otimes \rho_{F}^{(2)}\hspace*{-1mm},
\end{eqnarray*}
\noindent 
so it follows that the state does not factorize for any $t\geq 0$. \hfill{\ \rule{0.5em}{0.5em}} %\qed

\vspace*{1mm}

%
%   MAIN ATTRACTIVITY THEOREM
%

Note that proper initialization plays a more critical role in the DFS-
than in the NS-context, as a consequence of Proposition \ref{NSlink}.
If $\Hi_F\not ={\mathbb C}$ and the initial state is not factorized on
the $NSF$-block, the reduced state on $\Hi_{NS}$ still evolves
unitarily, provided that the generator satisfies the conditions given
in Corollary \ref{NScondition}.  Practically, this means that there is
no actual need to require a (factorized) subsystem-initialized state,
as long as a bounded error on the NS-component can be tolerated.  For
an imperfectly-initialized DFS, unitary evolution on the intended
block can only occur if $L_{P,k}\equiv 0$, making attractivity a
compelling option if the latter does not hold.  Accordingly, our main
emphasis is on attractivity in the DFS-case, which may be guaranteed
by invoking a specialization of the Krasowskii-LaSalle invariance
principle (see e.g.~\cite{khalil-nonlinear}).

\begin{thm}[{Attractive Subspace}]\label{attractionsubspace} 
Let $\Hi_I=\Hi_{S}\oplus\Hi_R$ ($\Hi_F=\C$), and let $\Hi_{S}$ support an
invariant subspace under ${\cal L}$. Assume that there exists a
continuously differentiable function $V(\rho)\geq 0$ on ${\mathfrak
D}(\Hi_{I}),$ such that $\dot V(\rho)\leq 0$ on imperfectly
initialized states in ${\mathfrak D}(\Hi_{I})\setminus{\mathfrak
I}_S(\Hi_{I})$. Let
\begin{eqnarray*}
{\cal W}&=&\{\rho\in{\mathfrak D}(\Hi_{I})|\,\dot V(\rho)= 0\},\\
{\cal Z}&=&\{\rho\in{\mathfrak D}(\Hi_{I})|\, {{\trace}}[\Pi_R{\cal
L}(\rho)]= 0\},
\end{eqnarray*}
where $\Pi_R$ is the orthogonal projector on $\Hi_R$. If ${\cal
W}\cap{\cal Z}\subseteq{\mathfrak I}_{S}(\Hi_{I})$, then $\Hi_S$ is
attractive.\end{thm}
\proofn Consider
$V_1(\rho)=\trace(\Pi_R\rho)+\trace(\Pi_R\rho)V(\rho).$ It is zero
iff $\rho_R=0$, i.e. for perfectly initialized states. 
Computing ${\cal L}(\rho)$, we get for the $R$ block: 
\begin{eqnarray*}
%\label{eq:LR} 
\begin{split} {\cal L}(\rho)_{R}=&-i[H_{R},\rho_{R}]+i\rho_P^\dag H_P-iH_P^\dag\rho_P \\&
-\frac{1}{2}\sum_{k}\left(
2L_{R,k}\rho_{R}L^\dag_{R,k}-\{L_{R,k}^\dag
L_{R,k},\,\rho_{R}\}\right. \\&
+L_{Q,k},\rho_{SF}L_{Q,k}^\dag-2\{L_{P,k}^\dag L_{P,k},\rho_R\}
\\ &+2L_{R,k}\rho_P^\dag L_{Q,k}^\dag 
+2L_{Q,k}\rho_P
L_{R,k}^\dag\\&-\rho_P^\dag(L_{SF,k}^\dag L_{P,k}-L_{Q,k}^\dag L_{R,k})\\ 
&\left. -(L_{P,k}^\dag L_{SF,k}+L_{R,k}^\dag L_{Q,k})\rho_P\right).
\end{split}
\end{eqnarray*}
\noindent 
Therefore, 
\beq\label{eq:pumping}\begin{split}\trace[\Pi_R{\cal
L}(\rho)]&= -\frac{1}{2}\trace\Big(\Big\{\sum_kL_{P,k}^\dag
L_{P,k},\rho_R\Big\}\Big)\\ &
=-\trace\Big(\sum_kL_{P,k}^\dag
L_{P,k}\rho_R\Big),\end{split}\eeq 
\noindent
is always negative or zero.  Hence
$$ \dot V_1(\rho)=\trace(\Pi_R{\cal
L}(\rho))(1+V(\rho))+\trace(\Pi_R\rho)\dot V(\rho)\leq 0,$$
\noindent 
for every ${\mathfrak D}(\Hi_{I}),$ and it is zero only in ${\cal
W}\cap{\cal Z}\cup{\mathfrak I}_S(\Hi_I)$.  If ${\cal W}\cap{\cal
Z}\subseteq {\mathfrak I}_S(\Hi_{I})$, by applying Krasowskii-LaSalle
invariance theorem, we conclude. \hfill{\ \rule{0.5em}{0.5em}} %\qed

\vspace*{1mm}

\noindent The following result immediately follows:

\begin{cor}\label{fullrank} Assume that 
$\Hi_I=\Hi_{S}\oplus\Hi_R$ ($\Hi_F=\C$), and let $\Hi_{S}$ support an
invariant subspace under ${\cal L}$. Assume that
\beq\label{eq:fullrank}\sum_kL_{P,k}^\dag L_{P,k}>0,\eeq
\noindent
where $>$ means strictly positive. Then $\Hi_S$ is
attractive.\end{cor}

\proofn It suffices to note that \eqref{eq:fullrank} guarantees that
\eqref{eq:pumping} in the proof of the Theorem above is zero iff
$\rho_R=0.$ The conclusion follows by taking a $V(\rho)$ constant and
positive on ${\mathfrak D}(\Hi_{I}).$\hfill{\ \rule{0.5em}{0.5em}}
%\qed

\vspace*{1mm}

{\em Remark:} From considerations on the rank of the l.h.s. of
\eqref{eq:fullrank} and the $n\times r$ dimension of $L_{P,k},$ the
condition of Corollary \ref{fullrank} may be obeyed only if $n\geq r$,
i.e. $\dim(\Hi_{S})\geq\dim(\Hi_{R}).$ An application of this result
will be given in Section \ref{control}.

\vspace*{1mm}

Proposition \ref{attractionfactor} and Theorem
\ref{attractionsubspace} (or Corollary \ref{fullrank}), may be
combined in order to obtain {\em sufficient conditions} for
attractivity in the general NS case.

\begin{prop} 
Let $\Hi_I=(\Hi_{S}\otimes\Hi_F)\oplus\Hi_R$, and let $\Hi_{S}$ be
an invariant subsystem under ${\cal L}$, with $$\bar{\cal
L}_F(\cdot)=\trace_S( \Pi_{SF}{\cal L}(\cdot)\Pi_{SF}).$$
\noindent
Assume that there exist a continuously differentiable functional
$V(\rho)\geq 0$ on ${\mathfrak D}(\Hi_{I}),$ such that $\dot
V(\rho)\leq 0$ on imperfectly initialized states in ${\mathfrak
D}(\Hi_{I})\setminus{\mathfrak I}_{SF}(\Hi_{I})$. Let ${\cal W},{\cal
Z}$ be defined as in Theorem \ref{attractionsubspace}. If $\bar{\cal
L}_F(\cdot)$ is relaxing and ${\cal W}\cap{\cal Z}\subseteq{\mathfrak
I}_{SF}(\Hi_{I})$, then $\Hi_S$ is attractive.\end{prop}

\proofn From Corollary \ref{invariancecondition}, $\bar{\cal
L}_F(\cdot)$ is a Markovian generator on $\Hi_F,$ thus it makes sense
to require that it is relaxing. Let $\hat\rho_{F}$ be its unique
attractive state. Observe that from Theorem \ref{attractionsubspace},
the state will asymptotically have support only on
$\Hi_{S}\otimes\Hi_F.$ Let $\bar\rho_{SF}(t)=\Pi_{SF}\rho(t)
\Pi_{SF}^\dag$: By defining $\tau=t/2,\, s=t/2$, and by invoking the
Markovian property, along with the above observation, invariance, and
Proposition \ref{attractionfactor} we may write:
\begin{eqnarray*}
\lim_{t\rightarrow\infty}\bar \rho_{SF} (t) & = &
\lim_{\tau,s\rightarrow\infty}\Pi_{SF}{\cal T}^\tau (\rho(s))\Pi_{SF}^\dag\\&=&
\lim_{\tau\rightarrow\infty}{\cal T}^\tau_{S}\otimes {\cal T}^\tau_{F}
(\lim_{s\rightarrow \infty} \bar\rho_{SF}(s))\\&=&
\lim_{\tau\rightarrow\infty}{\cal T}^\tau_{S}
(\lim_{s\rightarrow\infty}\bar\rho_{NS}(s))\otimes
\hat\rho_{F}.
\end{eqnarray*}
\noindent
The last equality comes from the fact that $\lim_{s\rightarrow \infty}
\bar\rho_{SF}(s)$ is certainly bounded, and can be always written in
the form $\lim_{s\rightarrow \infty} \sum_kP_{S,k}(s)\otimes
\rho_{F,k}$ by choosing a basis of ${\frak B}(\Hi_F)$ of density
operators $\{\rho_{F,k}\}$ and by following ideas similar to those in
the proof of Proposition \ref{attractionfactor}.\hfill{\
\rule{0.5em}{0.5em}}

\vspace*{-1mm} 

\section{Control Applications}

\subsection{Quantum trajectories and Markovian output feedback}
\label{markoviancontrol}

Building on pioneering work by Belavkin \cite{belavkin-towards}, it
has been long acknowledged for a diverse class of controlled quantum
system that intercepting and feeding back the information leaking out
of the system allow to better accomplish a number of desired control
tasks (see
\cite{wiseman-milburn,wiseman-feedback,doherty-linear,mabuchi-science,ticozzi-feedbackDD,
ganesan-feedback} for representative contributions).  This requires
the ability to both effectively monitor the environment and control
the target evolution through time-dependent Hamiltonian perturbations
which depend upon the measurement record.  We begin by recalling some
well-established continuous-measurement models for the conditioned
dynamics, as originally developed in the quantum-optics setting by
Wiseman and Milburn \cite{wiseman-milburn,wiseman-feedback}.

The basic setting is a measurement scheme which mimicks optical
homo-dyne detection for field-quadrature measurements, whereby the
target system (e.g. an atomic cloud trapped in an optical cavity) is
indirectly monitored via measurements of the outgoing laser field
quadrature \cite{wiseman-milburn,thomsen}.  Let the measurement record
be denoted by $Y_t$ (e.g. a photo-current in the above setting), and
let $( \Omega, {\cal E}, P) $ a (classical) probability space with an
associated $ \{ W_t , t\in \mathbb{R}^+ \} $ standard
$\mathbb{R}$-valued Wiener process. The homo-dyne detection
measurement record may then be written as the output of a stochastic
dynamical system of the form: \beq
\label{photocurrent} dY_t=\eta\,\tr(M\rho_t+\rho_t
M^\dag)dt+\sqrt{\eta}dW_t, \eeq 
\noindent
where $\rho_t\equiv \rho(t)$ is the system state at time $t$, $M$ is
the measurement operator determining the system-probe interaction, and
$0\leq\eta\leq 1$ quantifies the efficiency of the measurement.

The real-time knowledge of the photo-current provides additional
information on the dynamics, leading to a {\em stochastic master
equation} (SME) for the conditional evolution:
\begin{eqnarray}
d\rho_t &\hspace*{-2mm}=\hspace*{-3mm} &({\cal F}( H,\rho_t) + {\cal
 D}(M, \rho_t)) dt + {\cal G} (M, \rho_t) d W_t \label{eq:SME} \\ &
 \hspace*{-2mm}= \hspace*{-3mm}&\Big( -i[H,\rho_t]+\eta(M\rho_t M^\dag
 -\frac{1}{2}\{M^\dag M,\,\rho_t\})\Big)dt \nonumber 
 \\ & &+
 \sqrt{\eta }\Big[M\rho_t+\rho_t
 M^\dag- \tr( M\rho_t+\rho_t M^\dag) \rho_t \Big]dW_t.  \nonumber
\end{eqnarray}
Here, ${\cal F}$ is the Hamiltonian generator, whereas ${\cal
D}(M,\rho_t)$ and ${\cal G}(M,\rho_t)$ 
%=\sqrt{\eta}(M\rho_t+\rho_tM^\dag-\tr(M\rho_t+\rho_tM^\dag)\rho_t) $
are the Lindblad noise channel and the ``diffusion'' contribution due
to the weak measurement of $M$. Given an
initial condition $\rho_0$, the solution $\rho_t$
exists, is confined to ${\mathfrak D}(\Hi_I)$, and is adapted to the
filtration induced by $ \{ W_t , t\in \mathbb{R}^+ \} $(see e.g.
\cite{belavkin-filtering,vanhandel-feedback}).  As the SME
\eqref{eq:SME} is an It$\hat{\mbox{o}}$ stochastic differential
equation, to obtain the average evolution generator it suffices to
drop the martingale part ${\cal G}(M,\rho_t)dW_t$.  Thus, the
``unconditional'' evolution obeys a deterministic QDS generator of the
form \eqref{eq:lindblad}.  Notice that the diffusion term plays the
role of the innovation part of a nonlinear Kalman-Bucy filter, and
that the conditional state follows continuous trajectories, whereby
the name of {\em quantum trajectories} approach in the quantum-optics
literature \cite{carmicheal}.

In what follows, we assume perfect detection, that is, $\eta=1$,
unless otherwise specified (see \ref{comparison} and
\ref{sec:conclusions}).  In \cite{wiseman-feedback}, it has been
argued that the photo-current can be instantaneously fed back to
further modify the dynamics, still maintaining the Markovian character
of the evolution.  This motivates considering a {\em Hamiltonian
feedback superoperator} of the form \beq\label{eq:ill-fb} \ddt
\rho_t^{f}= \ddt Y_t {\cal F}(F,\rho_t), \;\; F=F^\dagger, \eeq
\noindent 
which is, however, ill-defined given the stochastic nature of $Y_t$. 
In order to obtain a feedback Markovian evolution, \eqref{eq:ill-fb}
has been interpreted as an \vv{implicit} Stratonovich stochastic differential
equation \cite{wiseman-feedback}. Its It$\hat{\mbox{o}}$ equivalent form is:
\beq\label{eq:ito-fb}d\rho_t^{f}={\cal
F}(F,\rho_t)dY_t+\frac{1}{2}{\cal F}^2(F,\rho_t)dt.\eeq

\noindent 
Thus, one can consider the infinitesimal evolution resulting from the
feedback followed by the measurement action $
\rho_t+d\rho_t={\cal T^{F}}_{dt} \circ{\cal T^{M}}_{dt}(\rho_t),$ where:
$${\cal T^{F}}_{dt}(\rho_t)=\rho_t+{\cal F}(F,\rho_t)dY_t+\frac{1}{2}{\cal
F}^2(F,\rho_t)dt,$$ 
%and: 
$${\cal T^{M}}_{dt}(\rho_t)=\rho_t+({\cal
F}( H,\rho_t) + {\cal D}(M, \rho_t)) dt  + {\cal G} (M, \rho_t) d
W_t.$$
\noindent 
Substituting the definitions and using It$\hat{\mbox{o}}$'s rule, 
it yields:
\beq\label{eq:markovianSME} 
\begin{split}d\rho_t & = \Big({\cal F}(
H,\rho_t)+ {\cal D}(M, \rho_t)+{\cal F}(
F,M\rho_t+\rho_tM^\dag)\\&
+\frac{1}{2}{\cal F}^2(F,\rho_t)\Big)dt+
\Big({\cal G} (M, \rho_t)+{\cal F}(F,\rho_t) \Big)d W_t.
\end{split}\eeq

\noindent 
Dropping again the martingale part and rearranging the remaining terms
leads to the Wiseman-Milburn {\em Markovian Feedback Master equation}
(FME) \cite{wiseman-milburn,wiseman-feedback}: \beq\label{eq:MME} \ddt
\rho_t={\cal F}\Big(H+ \frac{1}{2}(FM+M^\dag F),\rho_t \Big)+{\cal
D}(M - iF,\rho_t).\eeq
\noindent
In the following sections, we will tackle state-stabilization and
NS-synthesis problems for controlled Markovian dynamics described by
FMEs.

\vspace*{-2mm} 

\subsection{Control assumptions}\label{control}

The feedback state-stabilization problem for Markovian dynamics has
been extensively studied for the single-qubit
case~\cite{wang-wiseman,wiseman-bayesian}. In particular, conditions
for achieving a pure steady state have been identified in
\cite{yamamoto-purestate}.  In the existing literature, however, the
standard approach to design a Markovian feedback strategy is to
specify {both} the measurement and feedback operators $M,F,$ and to
treat the measurement strength and the feedback gain as the relevant
control parameters accordingly.
Here we will pretend to have more
freedom, considering, for a fixed measurement operator $M$, {\em both}
$F$ and $H$ as tunable control Hamiltonians.

\vspace*{1mm}

\begin{defin} [CHC] A controlled FME of the form \eqref{eq:MME} 
supports {\em complete Hamiltonian control} (CHC) if (i) arbitrary
feedback Hamiltonians $F\in {\mathfrak H}(\Hi_I)$ may be enacted; (ii)
arbitrary {\em constant} control perturbations $H_c\in {\mathfrak
H}(\Hi_I)$ may be added to the free Hamiltonian $H$.
\end{defin}

\vspace*{1mm}

As we shall see, this leads to both new insights and constructive
control protocols for systems where the noise operator is a
generalized angular momentum-type observable, for generic
finite-dimensional systems.  While assuming that the implementation of
arbitrary coherent Hamiltonians poses no problem is in line with
standard universality constructions for open quantum systems
\cite{lloyd-engineering,altafini-open}, from a physical standpoint the
CHC assumption is certainly demanding and should, as such, be
carefully scrutinized on a case by case basis. In particular,
constraints on the allowed Hamiltonian contributions relative to the
Lindblad dissipator may emerge, notably in so-called weak-coupling
limit derivations of Markovian models~\cite{alicki-lendi}.  
A first,
interesting consequence of assuming CHC emerges directly from the
following observation:

\vspace*{1mm}

\begin{lemma}\label{tracechange} The Markovian generator
\beq\label{eq:gen1} \ddt \rho_t=-i[H,\,\rho_t]+\sum_k{\cal
D}(L_k,\rho_t)\eeq 
is equivalent to 
\beq\label{eq:gen2} \ddt
\rho_t=-i[H+H_c,\,\rho_t]+\sum_k{\cal D}(\tilde L_k,\rho_t),\eeq
where for all $k,$ and $c_k\in\C$: 
\beq
\label{eq:Ltrans}
\tilde L_k=L_k+c_k {\mathbb I},\quad
 H_c=- i\sum_k (c_k^*L_k-c_kL_k^\dag).\eeq
\end{lemma}
\proofn Consider $k=1$: 
\bea {\cal D}(\tilde L,\rho)&=& \tilde
L\rho\tilde L^\dag-\frac{1}{2}\{\tilde L^\dag \tilde
L,\,\rho\}\nonumber\\ &=& L\rho L^\dag-\frac{1}{2}\{L^\dag
L,\,\rho\}+[c^*L-cL^\dag,\,\rho] \nonumber\\ &=&
-i[(ic^*L-icL^\dag),\,\rho]+{\cal D}(L,\rho).
\label{eq:tracechange}
\eea 
\noindent
Notice that $i(c^*L-cL^\dag)$ is Hermitian. For $k>1$, it suffices to
add up the correction parts in \eqref{eq:tracechange} for different
$k$'s, and use the linearity of the commutator.  \hfill{\
\rule{0.5em}{0.5em}} %\qed

\vspace*{1mm}

\noindent 
Note that for Hermitian $L$ and real $c$, $H_c=0$.  In general, by
exploiting CHC, we may vary the trace of the Lindblad operators
through transformations of the form
\eqref{eq:Ltrans}\footnote{Interestingly, this corresponds, in
physical terms, to a variation of the local oscillator in the optical
homo-dyne detection setting \cite{wiseman-feedback}.}, and, if needed
or useful, appropriately counteract the Hamiltonian correction $H_c$
with a {\em constant} control Hamiltonian.  This may allow to
stabilize subsystems that are not invariant for the uncontrolled
equation, {\em without directly modifying the non-unitary part}. In
addition to this, restricting to such open-loop, constant control
Hamiltonians avoids additional difficulties which are related to
reconcile the Markovian limit with generic time-varying
perturbations~\cite{alicki-lendi,lendi-timedependent}.

\vspace*{1mm}

{\em Example 3.} Consider a generator of the form:
$$\ddt \rho(t)=-{i}[\sigma_z,\rho(t)]+\Big(L \rho(t) L^\dag-
\frac{1}{2}\{L^\dag L,\,\rho(t)\}\Big),$$
\noindent
where $L=\sigma_z+\sigma_+$. Suppose that the task is to make
$\rho_d=\textrm{diag}(1,0)$ invariant. Since $H_P=0,L_S=1,L_p=1,$
invariance is not ensured by the uncontrolled dynamics. Using the
above result, it suffices to apply a constant Hamiltonian
$H_c=-i(L-L^\dag)=\sigma_y$.  The desired state turns out to be
also attractive, see Proposition \ref{stable2level} below.

\vspace*{-2mm} 

\subsection{Pure-state preparation with Markovian feedback: 
Two-level systems}\label{2level}

As mentioned, the problem of stabilizing an arbitrary pure state for a
two-level atom is discussed in detail in \cite{wang-wiseman} for
$H=\alpha\sigma_y,$ $M=\sqrt{\gamma}\sigma_-$,
%=\sqrt{\gamma}/2(\sigma_x-i\sigma_y)$, (the lowering operator), 
and $F=\lambda\sigma_y$, with the Hamiltonian, measurement, and
feedback strength parameters ( $\alpha\,,\,\gamma,\,\lambda,$
respectively), treated as the control design parameters.  In terms of
a standard Bloch sphere parametrization of the state set,
$\rho=1/2\mathbb{I}_2+1/2(x\sigma_x+y\sigma_y+z\sigma_z),$ with
$0\leq|(x,y,z)|\leq 1$, it is proved there that any pure state in the
$xz$ plane can be made invariant and attractive, with the only
exception of the states on the equator of the sphere. The possibility
of relaxing the perfect detection assumption is also addressed.

Our perspective differs not only because we mainly focus on continuous
measurement of {\em Hermitian spin observables}, but more importantly
because we start from identifying what constraints must be imposed to
a Lindblad equation for a two-dimensional system as in
\eqref{eq:lindblad} for ensuring that one of the system's pure states
is an attractive equilibrium. Without loss of generality, let such a
state be written as $\rho_d=\diag(1,0)$, and write, accordingly,
$$ L_k=\left(%
\begin{array}{cc}
  l_{k,S} & l_{k,P} \\
  l_{k,Q} & l_{k,R} \\
\end{array}%
\right),\quad H=\left(%
\begin{array}{cc}
  h_{S} & h_{P} \\
  h_{P}^* & h_{R} \\
\end{array}%
\right). $$

\begin{prop}\label{stable2level} The pure state $\rho_d=\diag(1,0)$ is a globally 
attractive, invariant state for a two-dimensional quantum system
evolving according to \eqref{eq:lindblad} iff:
\begin{eqnarray}&&i h_{P}-\frac{1}{2} \sum_k l_{k,S}^*l_{k,P}=0,\label{eq:c1}\\
&&l_{k,Q}=0,\quad\forall k, \label{eq:c2}\end{eqnarray}
and there exists a $\bar k$ such that $l_{\bar k,P}\neq 0$.
\end{prop}

\proofn Eqs. \eqref{eq:c1}-\eqref{eq:c2} imply the invariance
conditions of Corollary \ref{invariancecondition}, hence $\rho_d$ is
stable. For the choice $L^{D}_k=\mbox{diag}( l_{S,k}, l_{R,k})$,
%\left(%
%\begin{array}{cc}
%  l_{S,k} & 0 \\
%  0 & l_{R,k} \\
%\end{array}%
%\right),$ for each $k$,
every diagonal state would clearly be stationary (directly from the
form of ${\cal L}(\cdot)$, or by Proposition \ref{hermitianL}). Hence
it must be $l_{P,k}\neq 0$ for some $k$. To prove that $\rho_d$ is the
only attractive point for \eqref{eq:lindblad}, it suffices to note
that $l_{\bar k,P}\neq 0$ is the two-dimensional version of the
sufficient condition for attraction given in Corollary \ref{fullrank}.
\hfill{\ \rule{0.5em}{0.5em}}

\vspace*{1mm}

{\em Remark:} Observe that, even if $\rho_d$ is stable for the
unconditional, {\em averaged} dynamics over the trajectories of
\eqref{eq:SME}, because it is pure it cannot be obtained as a convex
combination of other states. Thus, $\rho_d$ must be the {\em
asymptotic limit of each trajectory with probability one}.  In fact,
any invariant set different from $\rho_d$ alone could not have it as
average. We provide next a characterization of the stabilizable
manifold.

\vspace*{1mm}

\begin{prop} 
Assume CHC. For any measurement operator $M$, there exist a feedback
Hamiltonian $F$ and a Hamiltonian compensation $H_c$ able to stabilize
an arbitrary desired pure state $\rho_d$ for the FME \eqref{eq:MME}
iff \vspace*{-1mm}
\beq\label{stabilize2}[\rho_d,(M+M^\dag)]
%[\rho_d,\mbox{Herm}(M)]
\neq0.\end{equation}
\end{prop} 
\vspace*{1mm}
\proofn Consider as before a basis where $\rho_d=\diag(1,0)$, and let
$M^H$ and $M^A$ denote the Hermitian and anti-Hermitian part of $M$,
respectively.  By \eqref{stabilize2}, $M^H$ cannot be diagonal in the
chosen basis.  In fact, assume $M^H$ to be diagonal, then, by
Proposition \ref{stable2level}, $M^S-F$ must be brought to diagonal
form to ensure invariance of $\rho_d$. Hence, by the same result, it
follows that $\rho_d$ cannot be made attractive. However, if $M^H$ is
not diagonal, we can always find an appropriate $F$ in order to get an
upper diagonal $L=M^H+i(M^S-F)$, and $H'=H+(FM+M^\dagger F)/2$. To
conclude, it suffices to devise a compensation Hamiltonian $H_c$ such
that the condition $i (H'+H_c)_{P}-\frac{1}{2}l_{S}^*l_{P}=0$ is
satisfied.  \hfill{\ \rule{0.5em}{0.5em}}

\vspace*{1mm}

The above proof naturally suggests a constructive algorithm for
designing the feedback and correction Hamiltonian required to
stabilize the desired state. From our analysis, we also recover the
results of \cite{wang-wiseman} recalled before.  For example, the
states that are never stabilizable within the control assumptions of
\cite{wang-wiseman} are the ones commuting with the Hermitian part of
$M=\sigma_+,$ that is, $M^H=\sigma_x.$ On the $xz$ plane in the
Bloch's representation, the latter correspond precisely to the
equatorial points. The following example serves to illustrate the
basic ideas we shall extend to the $d$-level case.

\noindent
{\em Example 4:} The simplest choice to obtain an attractive generator
is to engineer a dissipative part determined by
$L=\sigma_+=\left(%
\begin{smallmatrix}
  0 & 1 \\
  0 & 0 \\
\end{smallmatrix}
\right).$ Let $H=n_0 \mathbb I_2 + n_x\sigma_x + n_y\sigma_y + n_z\sigma_z$,
with $n_0,n_x,n_y, n_z \in \R$.  Consider e.g. $M=\frac{1}{2}\sigma_x$
and $F=-\frac{1}{2}\sigma_y$. Notice that in this case
$\frac{1}{2}(FM+M^\dag F)=0$, thus $H'=H$.  Substituting in the FME
\eqref{eq:MME}, one clearly obtain the desired result, provided that
$H_c=-n_x\sigma_x -n_y \sigma_y$.

\vspace*{1mm}

The spin measurement models considered above have been already
exploited for stabilization problems (see
e.g. \cite{vanhandel-feedback}), although in the context of strategies
necessitating a real-time estimate of the state \cite{doherty-linear}
-- so-called {\em Bayesian feedback} techniques in the physics
literature~\cite{wiseman-bayesian}.  Assume that it is possible to
continuously monitor a single observable, e.g. $\sigma_x$ in the above
example. Since the choice of the reference frame for the spin axis is
conventional, by suitably adjusting the relative orientation of the
measurement apparatus and the sample, it is then in principle possible
to prepare and stabilize any desired pure state with the same control
strategy.

\vspace*{-6mm} 

\subsection{Extension to multi-level systems}

The previous two-dimensional results naturally extend to generic
$d$-level systems.  This will also provide an example of an attractive
state, which does not satisfy the sufficient condition of Proposition
\ref{fullrank}.  Let the pure state to be FME-stabilized be written as
$\rho_d=\diag(1,0,\ldots, 0)$. Under CHC, we may without loss of
generality assume $H$ to be diagonal in this basis. 

\vspace*{1mm}

\begin{prop} The pure state $\rho_d$ is a globally attractive, 
invariant state for the FME \eqref{eq:MME} conditioned over the
continuous measurement of the operator: $$
M=\frac{1}{2}\left(\begin{array}{cccc} 
0 & m_1 &   & 0 \\ 
m_1 & 0 & \ddots & \\ 
 & \ddots &\ddots &m_{d-1}\\ 
0 &   & m_{d-1}& 0  
\end{array}
\right),$$ and a Markovian feedback Hamiltonian: $$
F=\frac{i}{2}\left(\begin{array}{cccc} 
0 & m_1 &   & 0 \\ 
-m_1 & 0 & \ddots & \\ 
 & \ddots &\ddots &m_{d-1}\\ 
0 &   & -m_{d-1}& 0  
\end{array}
\right),$$
with $m_i\neq 0,$ for $i=1,\dots, (d-1)$.
\end{prop}

\proofn First, observe that $L=(l_{i,j})=M-iF%=2 \left(\begin{array}{ccccc}
%  0 & m_1 & 0 &  \cdots & 0 \\
%  0 & 0 & \ddots & \ddots &\vdots  \\
%  \vdots &\ddots & \ddots & \ddots & 0 \\
%   &  &\ddots &\ddots & m_{d-1}\\
%  0 & \cdots & &0& 0 \\
%\end{array}
%\right).$$\vspace*{-1mm}
,$ the only elements different from zero are $l_{i,i+1}=m_i.$
 By writing $\rho=(\rho_{ij})_{i,j=1,\ldots,d}$, one gets:
\vspace*{2mm}
\bea\label{eq:proof2}&&\hspace{-8mm}{\cal D}(L,\rho)
= 4\left(
\begin{array}{cccc}
  m_1^*m_1\rho_{22} & \cdots & m_{d-1}^*m_{1}\rho_{2d} & 0 \\
  \vdots & \ddots &  & \vdots \\
  m_1^*m_{d-1}\rho_{d2} & \cdots & m_{d-1}^*m_{d-1}\rho_{dd} & 0 \\
  0 & \cdots & 0 & 0 \\
\end{array}%
\right)\nonumber-\vspace*{2mm}\\
&& \hspace{-9mm} 2\hspace*{-1mm}\left(\hspace*{-2mm}%
\begin{array}{cccc}
  0 & \hspace*{-2mm}|m_1|^2\rho_{12} & \hspace*{-2.5mm}\cdots & \hspace*{-3mm}|m_{d-1}|^2\rho_{1d} \vspace*{2mm}\\
  |m_1|^2\rho_{21} & \hspace*{-2.5mm}2|m_1|^2\rho_{22} & \hspace*{-2mm}\cdots & \hspace*{-3mm}(|m_1|^2\hspace*{-1mm}+\hspace*{-1mm}|m_{d-1}|^2)\rho_{2d} \\
  \vdots & \hspace*{-2.5mm}\vdots & \hspace*{-2mm}\ddots & \hspace*{-3mm}\vdots \vspace*{2mm}\\
  |m_{d-1}|^2\rho_{d1} &  \hspace*{-2.5mm}(|m_1|^2\hspace*{-1mm}+\hspace*{-1mm}|m_{d-1}|^2)\rho_{d2} & \hspace*{-2mm}\cdots & \hspace*{-3mm} 2|m_{d-1}|^2\rho_{dd} \\
\end{array}\hspace*{-3mm}
\right)\hspace*{-1mm}.\nonumber\\\nonumber\\
\eea
%\vspace*{-4mm}
%\bea\label{eq:proof2}&&\hspace{-8mm}{\cal D}(L,\rho) %=\nonumber\\ && 
%= 4\left(%
%\begin{array}{cccc}
%  m_1^*m_1\rho_{22} & \cdots & m_{d-1}^*m_{1}\rho_{2d} & 0 \\
%  \vdots & \ddots &  & \vdots \\
%  m_1^*m_{d-1}\rho_{d2} & \cdots & m_{d-1}^*m_{d-1}\rho_{dd} & 0 \\
%  0 & \cdots & 0 & 0 \\
%\end{array}%
%\right)\nonumber\\
%&&\hspace{-5mm}-2\left(%
%\begin{array}{cc}
%  0 & |m_1|^2\rho_{12}  \\
%  |m_1|^2\rho_{21} & 2|m_1|^2\rho_{22}  \\
%  \vdots & \vdots \\
%  |m_{d-1}|^2\rho_{d1} &  (|m_1|^2+|m_{d-1}|^2)\rho_{d2}  \\
%\end{array}\right.\nonumber\\
%&&\left.
%\begin{array}{cc}
%  \cdots & |m_{d-1}|^2\rho_{1d} \\
%   \cdots & (|m_1|^2+|m_{d-1}|^2)\rho_{2d} \\
%  \ddots & \vdots \\
%   \cdots &  2|m_{d-1}|^2\rho_{dd} \\
%   \end{array}
%\right).
%\eea 

\vspace*{-2mm}

\noindent 
Define $\hat{H}=\diag[0,1,\ldots,(d-1)]$. Thus, the function
$V_d(\rho)=\tr(\hat{H}\rho)$ is a valid global Lyapunov function for
the target state $\rho_d$ in ${\mathfrak D}(\Hi_I)$
\cite{khalil-nonlinear}.  Indeed, $V(\rho)\geq 0$ and $V(\rho)= 0$ iff
$\rho=\rho_d$. Computing $\dot V_d(\rho)=\tr(\hat{H}{\cal D}(L,\rho))$ using \eqref{eq:proof2}, one
obtains:
\[\begin{split}
\dot V_d(\rho)
=&4\bigg[\sum_{i=2}^{d-1} (i-1) |m_{ii}|^2\rho_{i+1,i+1}
-\sum_{j=1}^{d-1} j |m_{jj}|^2\rho_{j+1,j+1}\bigg]\\
&-4\sum_{i=1}^{d-1} |m_{ii}|^2\rho_{i+1,i+1}.\end{split}\]
\noindent
We conclude by applying Lyapunov stability theorem. 
Since $|m_{ii}|^2>0$ and $\rho_{i,i}\geq0$, the derivative is always
non-positive and can be zero {iff} $\rho_{i,i}=0,\,i=2,\ldots ,d,$
i.e. $\rho=\rho_d.$ \hfill{\ \rule{0.5em}{0.5em}} 

%\vspace*{1mm}

The matrix $\hat{H}$ in the above proof is essentially a Hamiltonian
with energy gaps renormalized to one, whereas $F$ and $M$ play a role
analogous to the $\sigma_y$ and $\sigma_x$ observables of the $d=2$
case.  Notice that their form is not different from that of standard,
higher-dimensional spin observables. %\cite{merzbacher}.

\vspace*{-2mm} 

\subsection{On Markovian-feedback state preparation}
\label{comparison}

The feedback strategies we consider preserve the Markovian character
of the open-system evolution.  Thus, in a sense, the corresponding
control problem may then be seen as a \vv{Markovian environment
design} problem \cite{wang-wiseman} -- implying that we may write the
QDS generator {\em independently} of the system state. This, along
with the remark in Section \ref{2level}, ensures the desired
convergence feature for {\em each} initial state (in other words,
robustness with respect to errors in the initial state estimation is
guaranteed).

The main advantage with respect to other feedback-design strategies is
represented by the potential ease in practical implementations, since
virtually no {signal-processing} stage is required in the realization
of the feedback loop. This should be contrasted with Bayesian feedback
strategies~\cite{vanhandel-feedback,wiseman-bayesian}, whereby an
updated state estimate has to be obtained through real-time
integration of \eqref{eq:SME}, and used to tailor a state-dependent
feedback action on the underlying evolution.  Such a task becomes
rapidly prohibitive as the dimensionality of the target system grows.

As a potential disadvantage, howewer, the Markovian output-feedback we
use requires strong control capabilities and perfect detection.  On
one hand, an infinite bandwidth is needed to feed back the measurement
output in real time.  On the other hand, both the feedback and
measurement parameters have to be accurately tuned, along with both
the system Hamiltonian and its control compensation, if
needed. Nevertheless, for state stabilization problems, one may assess
the role of the perfect-detection hypothesis and the possibility to
relax it.  If $\eta<1,$ the FME is modified as follows~\cite{thomsen}:
\bea\label{eq:imperfectMME0} 
\ddt \rho_t&\hspace*{-2mm}=\hspace*{-2mm}&{\cal F}\Big(H+1/2(FM+M^\dag
F),\rho_t \Big)
\nonumber \\&&+\hspace*{1mm}{\cal D}(M-iF,\rho_t)+\varepsilon \mathcal{D}(F,\rho_t),
\eea
\noindent
where we defined $\varepsilon= (1-\eta)/{\eta}$.

In \cite{alicki-lendi}, generators of the form
\eqref{eq:GKS}-\eqref{eq:lindblad} are rewritten in a convenient way
by choosing a suitable Hermitian basis in ${\frak B}(\Hi_i)\approx
\C^{d\times d}.$ In fact, endowing $\C^{d\times d}$ with the inner
product $\langle X,Y\rangle:=\trace(X^\dag Y)$ (Hilbert-Schmidt), we
may use a basis where the first element is
$\smallfrac{1}{\sqrt{d}}\mathbb{I}_d,$ and complete it with a
orthonormal set of Hermitian, traceless operators. This can always be
done for finite $d$, for example by employing the natural
$d$-dimensional extension of the Pauli matrices~\cite{alicki-lendi,
altafini-open}. In such a basis, all density operators are represented
by $d^2$-dimensional vectors
$\bar\rho=(\rho_0,\,\rho_1,\ldots,\,\rho_{d^2-1})^T,$ where the first
component $\rho_0$, relative to $\smallfrac{1}{\sqrt{d}}\mathbb{I}_d,$
is invariant and equal to $\smallfrac{1}{\sqrt{d}}$ for TP-dynamics.
Let $\rho_v=(\rho_1,\ldots,\,\rho_{d^2-1})^T.$ Hence, any QDS
generator ${\cal L}(\rho)$ must take the form: \beq \frac{d
}{dt}\bar\rho= \left(
\begin{array}{c|c}
  0 & 0    \\\hline
  C  &  D  
\end{array}
\right)\left(\begin{array}{c}
   1/\sqrt{d} \\
  \rho_v    
\end{array}
\right).  \eeq 
\noindent
Assume that the dynamics has a unique attractive state
$\bar\rho^{(0)}$.  Thus $D$ must be invertible and we obtain:
$$\bar\rho^{(0)}= \frac{1}{\sqrt{d}}\left(\begin{array}{c} 1\\
-D^{-1}C \end{array}\right).$$

Consider now a small perturbation of the generator depending on the
continuous parameter $\varepsilon,$ with $1-\delta<\eta<1,$ and
$\delta$ sufficiently small so that $(D+\varepsilon D')$ remains
invertible.  The generator becomes: 
\beq\frac{d \bar\rho}{dt}= \left[
\left(
\begin{array}{c|c}
  0 & 0    \\\hline
  C  &  D  
\end{array}
\right)
+\varepsilon\left(
\begin{array}{c|c}
  0 & 0    \\\hline
  C'  &  D'  
\end{array}
\right)
\right]
\left(\begin{array}{c}
   1/\sqrt{d} \\ 
\rho_v
\end{array}
\right),\eeq 
and the new attractive, unique equilibrium state is:
$$\bar\rho^{(\varepsilon)}=
\frac{1}{\sqrt{d}}\left(\begin{array}{c}1\\-(D+\varepsilon
D')^{-1}(C+\varepsilon C')\end{array}\right).$$ 
\noindent
Because $\bar\rho^{(\varepsilon)}$ is a continuous function of
$\varepsilon$, we are guaranteed that for a sufficiently high
detection efficiency the perturbed attractive state will be
arbitrarily close to the desired one in trace norm.  Therefore, if we
relax our control task to a state preparation problem with
sufficiently high fidelity,
%(or sufficiently low error probability), 
this may be accomplished with a sufficiently high detection
efficiency, yet strictly less than 1.

\vspace*{1mm}

Insofar as noise suppression is the intended task, we see how
monitoring a perfectly dissipative environment may be useful for
control process.  However, the ability to suppress the noise source
via feedback is necessarily limited by the form of \eqref{eq:MME}. It
is apparent that the feedback action is only able to {\em modify the
skew-Hermitian part of the noise operator $L$}, and even in cases
where this may suffice, (nearly) perfect detection is needed.
Nonetheless, Markovian feedback may prove to be extremely interesting
when only partial noise suppression is considered, for instance in
order to achieve longer coherence times.  In this spirit, we turn to
analyze how our techniques may be employed to synthesize DFSs or NSs
in the Markovian limit where open-loop control is not an option.

\vspace*{-2mm} 

\subsection{DFS synthesis with Markovian feedback}

As remarked, the feedback loop can only modify the skew-Hermitian part
of the measurement operator $M$, which imposes strict constraints on
the non-unitary generators that are able to be synthesized.  A natural
question is to what extent we might be able to generate DFSs or NSs by
closed-loop control. In the single-observable feedback setting under
examination, the DFS notion turns out to be appropriate.

\vspace*{1mm}

\begin{thm} \label{DFSgen}
Let $p=d/2,$ if $d$ is even, $p=(d+1)/2,$ if $d$ is odd, and assume
CHC for \eqref{eq:MME}.  Then a DFS of (at least) dimension $p$ can be
generated by Markovian feedback for every measurement operator $M$.
\end{thm}

\proofn A DFS for \eqref{eq:MME} can be generated, under CHC
hypotesis,  iff there exist a choice of basis such that:
\beqan
&&\hspace*{-7mm} L=M-iF= \left(
\begin{array}{c|c}
 c\mathbb{I}_{DFS} &  P   \\\hline
 0 &   R   \\  
\end{array}
\right)\\
&&\hspace*{-3mm}=\left(
\begin{array}{c|c}
 \textrm{Re}(c)\mathbb{I}_{DFS} &  P/2   \\\hline 
 P^\dag/2 &   R^H   \\  
\end{array}
\right)+i\left(
\begin{array}{c|c}
 \textrm{Im}(c)\mathbb{I}_{DFS} &  -iP/2   \\\hline
 iP^\dag/2 &   R^A   \\  
\end{array}
\right), \eeqan 
where we have decomposed $L$ into Hermitian (H) and skew-Hermitian
(A) parts as before.
The skew-Hermitian part can be arbitrarily modified under CHC
hypothesis, by choosing the appropriate $F$. Thus, it remains to prove
that there exists a basis where 
$M^H$ has the form of $L^H$ above, and the block proportional to the identity is (at least)
$p$-dimensional. $M^H$ is indeed Hermitian, and can be
diagonalized. Let $D^H$ be the diagonal matrix of the (real)
eigenvalues of $M^H.$ Then we are looking for a $U$ and a Hilbert space decomposition such that: 
\beq
\begin{split} 
&\hspace*{-2mm}\left( 
\begin{array}{c|c}
 \textrm{Real}(c)\mathbb{I}_{DFS} &  P/2   \\\hline
 P^\dag/2 &   R^H   \\  
\end{array}
\right)=
 UD^HU^\dag \\
&=\left(
\begin{array}{c|c}
 U_{DFS} &  U_P   \\\hline
U_Q &  U_R  \\  
\end{array}
\right)\hspace*{-1mm}\left(
\begin{array}{c|c}
D^H_{DFS} &  0  \\\hline
0 &  D^H_{R} \\  
\end{array}
\right)\hspace*{-1mm}\left(
\begin{array}{c|c}
 U_{DFS}^\dag &  U_Q^\dag   \\\hline
U_P^\dag &  U_R^\dag  \\ 
\end{array}
\right)
\end{split} \nonumber
\eeq

%\beq
%\begin{split} \hspace*{-2mm} UD^HU^\dag &\hspace*{-1mm}=\hspace*{-1mm}
%\left(
%\begin{array}{c|c}
% U_{DFS} &  U_P   \\\hline
%U_Q &  U_R  \\  
%\end{array}
%\right)\hspace*{-2mm}\left(
%\begin{array}{c|c}
%D^H_{DFS} &  0  \\\hline
%0 &  D^H_{R} \\  
%\end{array}
%\right)\hspace*{-2mm}\left(
%\begin{array}{c|c}
% U_{DFS}^\dag &  U_Q^\dag   \\\hline
%U_P^\dag &  U_R^\dag  \\ 
%\end{array}
%\right)\\
%&=
%\left(
%\begin{array}{c|c}
% \textrm{Real}(c)\mathbb{I}_{DFS} &  P/2   \\\hline
% P^\dag/2 &   R^H   \\  
%\end{array}
%\right).
%\end{split} \nonumber
%\eeq

\vspace{-5mm}

\noindent Hence, we want to find $p$ orthonormal vectors to stack in
$(U_{DFS}\,\,U_P)^\dag$ such that: \vspace*{-2mm}
$$ 
%\beq 
\left(
\begin{array}{c|c}
U_{DFS} &  U_P  \\
\end{array}
\right)\left(
\begin{array}{c|c}
D^H_{DFS} &  0  \\\hline
0 &  D^H_{R} \\  
\end{array}
\right)\left(
\begin{array}{c}
U_{DFS}^\dag\\ \hline U_P^\dag  
\end{array}
\right)=c'\mathbb{I}_{DFS},$$
%\eeq 
with $c'\in {\mathbb R}$. This is equivalent to ask that the
compression of $M^H$ to some subspace is equivalent to a scalar
matrix.  Let $u_1,\, u_2$ be normalized eigenvectors of $M^H$ of
eigenvalues $d_1,\, d_2$, respectively. Then by taking $u_3=\alpha
u_1+\beta u_2,$ such that $|\alpha|^2+|\beta|^2=1$, we can construct a
vector that is an eigenvector of $M^H$ restricted to the
one-dimensional subspace generated by $u_3$ itself. $u_3$ has then
eigenvalue $|\alpha|^2d_1+|\beta|^2d_2,$ a convex combination of the
former eigenvalues $d_1,\, d_2$. Take the diagonal elements of $D^H$
in descending order, and pair the first with the last, the second to
last but one, and so on. If $d$ is odd, the eigenvalue in the middle
will remain unpaired. Thus, from these pairs and the respective
eigenvectors we can then obtain $p$ new vectors as illustrated above,
which are all eigenvectors of $M^H$ restricted to their linear span,
with the same eigenvalue (in general, $c'$ will be a convex
combination of the two middle eigenvalues.  For odd $d$, it will be
the middle eigenvalue). From the resulting linear span, we can then
obtain the desired DFS, by choosing a feedback Hamiltonian $F$ such
that the $Q$ block of $M-i F$ is zero.  \hfill{\ \rule{0.5em}{0.5em}}
%\qed

%\vspace*{1mm}

{\em Remarks:} Note that the above proof provides a constructive {\em algorithm for generating the DFS}. The result is
potentially useful in light of ongoing efforts for efficiently finding
quantum information-preserving structures
\cite{knill-protected,kribs-prl,robin-ips}.  Both the CHC assumption and
the ability to perfectly monitor the noise channel are demanding for
present experimental capabilities, however the promise of a new
technique to generate DFSs may prompt further developments in this
direction.  In parallel, further study is needed in order to weaken
the above requirements, as it is likely to be possible in specific
contexts.

From another perspective, it is intriguing to compare Theorem
\ref{DFSgen} to the analysis of continuous-time quantum error
correction presented in \cite{ahn-feedback}.  In that case, the target
system is assumed to be a quantum register, and under the assumption
that {\em independent errors} are occurring on different qubits, a
Markovian feedback strategy is identified such that the closed-loop
behavior implements continuous-time quantum error correction for a
so-called ``stabilizer code'' \cite{nielsen-chuang}.  This may be seen
as equivalent to the generation of a DFS able, in particular, to
encode $(n-1)$ logical qubits in a $2^n$ dimensional space.  Even if
our analysis follows different lines, the setting for our
DFS-generation problem is similar, and our result consistently leads
to the same encoding efficiency for $d=2^n$ -- provided we can
compound the noise effect in a single measurement
operator. Interestingly, no assumption is made at this stage on the
structure of the Hilbert space, neither do we impose any constraint on
the form of the ``error'', that is, the measurement or noise operator
in our case.

Before concluding, we present a simple example of generation of
attractive DFS for coupled qubits via Markovian feedback, which
further illustrates some of our results.

\vspace*{.5mm}

{\em Example 4.}  Consider $\Hi=\Hi_q\otimes\Hi_q$,
$\Hi_q=\textrm{span}\{\ket{0},\ket{1}\}$, and a controlled closed-loop
evolution driven by \eqref{eq:MME}, with $H$ diagonal and
$M=\sigma_x\otimes\sigma_x$. Assume that we are able to monitor $M$
and actuate the feedback Hamiltonian $F=-\sigma_y\otimes\sigma_x.$
Then $L=M-iF=2\sigma_-\otimes \sigma_x$.
% with $\sigma=1/2(\sigma_x-i\sigma_y)$. 
If we consider
$\Hi_{DFS}=\textrm{span}\{\ket{0}\otimes\ket{0},\ket{0}\otimes\ket{1}\},$
we obtain block-decomposition of the form \eqref{eq:blocks}, where $L$
is such that $L_{DFS}=0,$ $L_Q=0,$ $L_P=\sigma_x$. Hence, by Corollary
\ref{NScondition} $\Hi_{DFS}$ is a DFS and by Corollary
\ref{fullrank}, we can prove it is attractive. Notice that the noise
operator $M=\sigma_x\otimes\sigma_x$ does admit noiseless subspaces,
e.g.
$\Hi'_{DFS}=\textrm{span}(\ket{+}\otimes\ket{+},\ket{+}\otimes\ket{-}),$
with $\ket{\pm}=1/\sqrt{2}(\ket{0}\pm\ket{1})$, but by Proposition
\ref{hermitianL}, none of them can be attractive. This shows how the
feedback-generated noiseless structure may offer an advantage with
respect to existing ones.

\vspace*{-2mm} 

\section{Discussion and Conclusion}\label{sec:conclusions}

We have revisited some fundamental concepts about Markovian dynamics
for quantum systems and restated the notion of a general quantum
subsystem in {\em linear-algebraic} terms.  A system-theoretic
characterization of invariant and noiseless subsystems for Markovian
quantum dynamical systems has been provided, with special attention on
key model-robustness issues relevant for practical applications.  In
particular, we have showed that, in order to avoid situations where
only fine-tuning of the Hamiltonian and dissipative terms would ensure
invariance of a given subsystem, condition \eqref{eq:nonrobust} must
be replaced by the independent conditions \eqref{eq:robust1} and
\eqref{eq:HamBlocks}.  This induces similar modifications on NS
invariance conditions.  This part of our work both puts on more
rigorous mathematical grounds and completes the existing literature on
the subject.

When imperfect subsystem initialization is considered, the conditions
to be imposed on the Markovian generator become more demanding, which
motivates the new notion of asymptotically stable {\em attractive}
subsystem. Interestingly, the possibility that the noise action may be
useful or necessary for remaining in the intended subsystem has been
independently considered before for specific QIP
settings~\cite{beige}. However, a formal generalization of the idea
and an analytical study were still missing.  Our linear-algebraic
approach, along with Lyapunov's techniques, provides explicit
stabilization results which have been illustrated in simple yet
paradigmatic examples.

In the second part of the work, the conditions identified for
subsystem invariance and attractivity serve as the starting point for
designing output-feedback Markovian strategies able to actively
achieve the intended quantum stabilization.  We have completely
characterized the state-stabilization problem for two-level systems
described by FMEs of the form~\eqref{eq:MME}.  While the analysis
assumed perfect detection efficiency, a perturbative argument
indicated how unique attractive states depend in a continuous fashion
on the model parameters.  Our suggested DFS generation strategy is
also crucially dependent on the perfect detection condition, as
otherwise the feedback-corrected FME would take the form
\eqref{eq:imperfectMME0}, implying an additional error component due
to finite efficiency.  Nonetheless, the norm of this portion of the
noise generator is bounded, and tends to zero for $\eta \rightarrow
1$. Therefore, even if the non-unitary dynamics cannot be counteracted
exactly, the time-scale of the residual noise action may still be
significantly reduced in the desired subspaces for sufficiently high
detection efficiency.

The Markovian, output-feedback techniques we employ have also been
compared, in terms of robustness features, with the Bayesian-feedback
approach. A key advantage of the Markovian approach lies in its
intrinsic design simplicity, which makes it possible to avoid a costy
real-time integration of the feedback-controlled master equation and
thereby paves the way to implementation in higher-dimensional systems.

Further work is needed in order to establish completely general
Markovian feedback stabilization results, including finite detection
efficiency and {\em multi-channel continuous monitoring}.  From an
algorithmic standpoint, it also appears worthwhile to investigate the
potential of the linear-algebraic approach in problems related to
finding NSs for a given generator, either under perfect or imperfect
knowledge.  Among the most interesting perspectives, additional
investigation is certainly required to establish the full power of
Hamiltonian control and Markovian feedback in generating NS
structures. This may point to new venues for producing protected
realizations of quantum information for physical systems whose
dynamics is described by quantum Markovian semigroups.

\vspace*{-2mm} 

\section{Acknowledgments}

It is a pleasure to thank Angelo Carollo for early discussions on DFS
conditions under Markovian dynamics, and Claudio Altafini, Augusto
Ferrante, Hideo Mabuchi, Michele Pavon, and Ramon van Handel for
valuable feedback.  F. T. acknowledges hospitality from the Physics
and Astronomy Department at Dartmouth College, under the support of the Constance and Walter Burke Special Projects Fund in Quantum Information Science.

\vspace{-5mm}

\bibliographystyle{IEEEtran}
%\bibliography{IEEEabrv,bibliography-2-1}

\bibliography{IEEEabrv,bibliography-2}

\begin{thebibliography}{10}
\providecommand{\url}[1]{#1}
\csname url@rmstyle\endcsname
\providecommand{\newblock}{\relax}
\providecommand{\bibinfo}[2]{#2}
\providecommand\BIBentrySTDinterwordspacing{\spaceskip=0pt\relax}
\providecommand\BIBentryALTinterwordstretchfactor{4}
\providecommand\BIBentryALTinterwordspacing{\spaceskip=\fontdimen2\font plus
\BIBentryALTinterwordstretchfactor\fontdimen3\font minus
  \fontdimen4\font\relax}
\providecommand\BIBforeignlanguage[2]{{%
\expandafter\ifx\csname l@#1\endcsname\relax
\typeout{** WARNING: IEEEtran.bst: No hyphenation pattern has been}%
\typeout{** loaded for the language `#1'. Using the pattern for}%
\typeout{** the default language instead.}%
\else
\language=\csname l@#1\endcsname
\fi
#2}}

\bibitem{sakurai}
J.~J. Sakurai, \emph{Modern Quantum Mechanics}.\hskip 1em plus 0.5em minus
  0.4em\relax Addison-Wesley, New York, 1994.

\bibitem{nielsen-chuang}
M.~A. Nielsen and I.~L. Chuang, \emph{Quantum Computation and
  Information}.\hskip 1em plus 0.5em minus 0.4em\relax Cambridge University
  Press, Cambridge, 2002.

\bibitem{viola-generalnoise}
E.~Knill, R.~Laflamme, and L.~Viola, ``Theory of quantum error correction for
  general noise,'' \emph{Physical Review Letters}, vol.~84, no.~11, pp.
  2525--2528, 2000.

\bibitem{knill-protected}
E.~Knill, ``On protected realization of quantum information,'' \emph{Physical
  Review A}, vol.~74, no.~4, pp. 042\,301:1--11, 2006.

\bibitem{zanardi-DFS}
P.~Zanardi and M.~Rasetti, ``Noiseless quantum codes,'' \emph{Physical Review
  Letters}, vol.~79, no.~17, pp. 3306--3309, 1997.

\bibitem{lidar-DFS}
D.~A. Lidar, I.~L. Chuang, and K.~B. Whaley, ``Decoherence-free subspaces for
  quantum computation,'' \emph{Physical Review Letters}, vol.~81, no.~12, pp.
  2594--2597, 1997.

\bibitem{viola-qubit}
L.~Viola, E.~Knill, and R.~Laflamme, ``Constructing qubit in physical
  systems,'' \emph{Journal of Physics A}, no.~34, pp. 7067--7079, 2001.

\bibitem{kempe-NS}
J.~Kempe, D.~Bacon, D.~A. Lidar, and K.~B. Whaley, ``Theory of decoherence-free
  fault-tolerant universal computation,'' \emph{Physical Review A}, vol.~63,
  no.~4, pp. 042\,307:1--29, 2001.

\bibitem{violaDygen}
L.~Viola, E.~Knill, and S.~Lloyd, ``Dynamical generation of noiseless quantum
  subsystems,'' \emph{Physical Review Letters}, vol.~85, no.~16, pp.
  3520--3523, 2000.

\bibitem{alicki-lendi}
R.~Alicki and K.~Lendi, \emph{Quantum Dynamical Semigroups and
  Applications}.\hskip 1em plus 0.5em minus 0.4em\relax Springer-Verlag,
  Berlin, 1987.

\bibitem{lidar-initializationDFS}
A.~Shabani and D.~A. Lidar, ``Theory of initialization-free decoherence-free
  subspaces and subsystems,'' \emph{Physical Review A}, vol.~72, no.~4, pp.
  042\,303:1--14, 2005.

\bibitem{doherty-linear}
A.~C. Doherty and K.~Jacobs, ``Feedback control of quantum systems using
  continuous state estimation,'' \emph{Physical Review A}, vol.~60, no.~4, pp.
  2700--2711, 1999.

\bibitem{davies}
E.~B. Davies, \emph{Quantum Theory of Open Systems}.\hskip 1em plus 0.5em minus
  0.4em\relax Academic Press, London, 1976.

\bibitem{petruccione}
H.~P. Breuer and F.~Petruccione, \emph{The Theory of Open Quantum
  Systems}.\hskip 1em plus 0.5em minus 0.4em\relax Oxford University Press,
  Oxford, 2002.

\bibitem{kraus}
K.~Kraus, \emph{States, Effects, and Operations: Fundamental Notions of Quantum
  Theory}, ser. Lecture Notes in Physics.\hskip 1em plus 0.5em minus
  0.4em\relax Springer-Verlag, Berlin, 1983.

\bibitem{lindblad}
G.~Lindblad, ``On the generators of quantum dynamical semigroups,''
  \emph{Communication in Mathematical Physics}, vol.~48, no.~2, pp. 119--130,
  1976.

\bibitem{qds}
V.~Gorini, A.~Frigerio, M.~Verri, A.~Kossakowski, and E.~C.~G. Sudarshan,
  ``Properties of quantum markovian master equations,'' \emph{Reports on
  Mathematical Physics}, vol.~13, no.~2, pp. 149--173, 1978.

\bibitem{gorini-k-s}
V.~Gorini, A.~Kossakowski, and E.~Sudarshan, ``Completely positive dynamical
  semigroups of n-level systems,'' \emph{Journal of Mathematical Physics},
  vol.~17, no.~5, pp. 821--825, 1976.

\bibitem{hille-phillips}
E.~Hille and R.~S. Phillips, \emph{Functional Analysis and Semigroups}.\hskip
  1em plus 0.5em minus 0.4em\relax American Mathematical Society, Providence,
  1957.

\bibitem{Boulant}
N.~Boulant, T.~F. Havel, M.~A. Pravia, and D.~G. Cory, ``Robust method for
  estimating the lindblad operators of a dissipative quantum process from
  measurements of the density operator at multiple time points,''
  \emph{Physical Review A}, vol.~67, no.~4, pp. 042\,322:1--12, 2003.

\bibitem{doyle-feedback}
J.~Doyle, A.~B. Francis, and A.~Tannenbaum, \emph{Feedback Control
  Theory}.\hskip 1em plus 0.5em minus 0.4em\relax MacMillan Publishing Company,
  New York, 1992.

\bibitem{ticozzi-robust}
F.~Ticozzi, A.~Ferrante, and M.~Pavon, ``Robust steering of n-level quantum
  systems,'' \emph{IEEE Transactions on Automatic Control}, vol.~49, no.~10,
  pp. 1742--1745, 2004.

\bibitem{kribs-prl}
D.~W. Kribs, R.~Laflamme, and D.~Poulin, ``Unified and generalized approach to
  quantum error correction,'' \emph{Physical Review Letters}, vol.~94, pp.
  180\,501:1--4, 2005.

\bibitem{verification}
L.~Viola and E.~Knill, ``Verification procedures for quantum noiseless
  subsystems,'' \emph{Physical Review A}, vol.~68, no.~3, pp. 032\,311:1--5,
  2003.

\bibitem{Klappenecker07}
A.~Klappenecker and K.~Sarvepalli, ``On subsystem codes beating the hamming or
  singleton bound,'' \emph{e-print arXiv.org:quant-ph/0703213}, 2007.

\bibitem{Kielpinski}
D.~Kielpinski, V.~Meyer, M.~A. Rowe, C.~A. Sackett, W.~M. Itano, C.~Monroe, and
  D.~J. Wineland, ``A decoherence-free quantum memory using trapped ions,''
  \emph{Science}, vol. 291, no. 5506, pp. 1013--1015, 2001.

\bibitem{violadfs}
E.~M. Fortunato, L.~Viola, J.~Hodges, G.~Teklemariam, and D.~G. Cory,
  ``Implementation of universal control on a decoherence-free qubit,''
  \emph{New J. Phys.}, vol.~4, p. 5.1, 2002.

\bibitem{nicolasQEC}
N.~Boulant, L.~Viola, E.~M. Fortunato, and D.~G. Cory, ``Experimental
  implementation of a concatenated quantum error-correcting code,''
  \emph{Physical Review Letters}, vol.~94, no.~13, pp. 130\,501:1--4, 2005.

\bibitem{viola-science}
L.~Viola, E.~M. Fortunato, M.~A. Pravia, E.~Knill, R.~Laflamme, and D.~G. Cory,
  ``Experimental realization of noiseless subsystems for quantum information
  processing,'' \emph{Science}, vol. 293, no. 5537, pp. 2059--2063, 2001.

\bibitem{nielsen-operatorschmidt}
M.~A. Nielsen, C.~M. Dawson, J.~L. Dodd, A.~Gilchrist, D.~Mortimer, T.~J.
  Osborne, M.~J. Bremner, A.~W. Harrow, and A.~Hines, ``Quantum dynamics as a
  physical resource,'' \emph{Physical Review A}, vol.~67, no.~5, pp.
  052\,301:1--19, 2003.

\bibitem{whaley-DFS}
R.~I. Karasik, K.-P. Marzlin, B.~C. Sanders, and K.~B. Whaley, ``Criteria for
  dynamically stable decoherence-free subspaces,'' \emph{e-print
  arXiv.org:quant-ph/0702243}, 2007.

\bibitem{spohn-approach}
H.~Spohn, ``Approach to equilibrium for completely positive dynamical
  semigroups of n-level system,'' \emph{Reports on Mathematical Physics},
  vol.~10, no.~2, pp. 189--194, 1976.

\bibitem{spohn-equilibrium}
------, ``An algebraic condition for the approach to equilibrium of an open
  n-level quantum system,'' \emph{Letters in Mathematical Physics}, vol.~2, pp.
  33--38, 1977.

\bibitem{khalil-nonlinear}
H.~K. Khalil, \emph{Nonlinear Systems}, 3rd~ed.\hskip 1em plus 0.5em minus
  0.4em\relax Prentice Hall, Upper Saddle River, New Jersey, 2002.

\bibitem{belavkin-towards}
V.~P. Belavkin, ``Theory of control of observable quantum systems,''
  \emph{Automatica and Remote Control}, vol.~44, pp. 178--188, 1983.

\bibitem{wiseman-milburn}
H.~M. Wiseman and G.~J. Milburn, ``Quantum theory of optical feedback via
  homodyne detection,'' \emph{Physical Review Letters}, vol.~70, no.~5, pp.
  548--551, 1993.

\bibitem{wiseman-feedback}
H.~M. Wiseman, ``Quantum theory of continuous feedback,'' \emph{Physical Review
  A}, vol.~49, no.~3, pp. 2133--2150, 1994.

\bibitem{mabuchi-science}
J.~M. Geremia, J.~K. Stockton, and H.~Mabuchi, ``Real-time quantum feedback
  control of atomic spin-squeezing,'' \emph{Science}, vol. 304, no. 5668, pp.
  270--273, 2004.

\bibitem{ticozzi-feedbackDD}
F.~Ticozzi and L.~Viola, ``Single-bit feedback and quantum dynamical
  decoupling,'' \emph{Physical Review A}, vol.~74, no.~5, pp. 052\,328:1--11,
  2006.

\bibitem{ganesan-feedback}
N.~Ganesan and T.~J. Tarn, ``Decoherence control in open quantum system via
  classical feedback,'' \emph{Physical Review A}, vol.~75, no.~3, pp.
  032\,323:1--19, 2007.

\bibitem{thomsen}
L.~K. Thomsen, S.~Mancini, and H.~M. Wiseman, ``Continuous quantum
  nondemolition feedback and unconditional atomic spin squeezing,''
  \emph{Journal of Physics B}, vol.~35, no.~23, pp. 4937--4952, 2002.

\bibitem{belavkin-filtering}
V.~P. Belavkin, ``Quantum stochastic calculus and quantum nonlinear
  filtering,'' \emph{Journal of Multivariate Analysis}, vol.~42, pp. 171--201,
  1992.

\bibitem{vanhandel-feedback}
R.~van Handel, J.~K. Stockton, and H.~Mabuchi, ``Feedback control of quantum
  state reduction,'' \emph{IEEE Transactions on Automatic Control}, vol.~50,
  no.~6, pp. 768--780, 2005.

\bibitem{carmicheal}
H.~Carmichael, \emph{An Open Systems Approach to Quantum Optics}, ser. Lecture
  Notes in Physics.\hskip 1em plus 0.5em minus 0.4em\relax Springer-Verlag,
  Berlin, 1993.

\bibitem{wang-wiseman}
J.~Wang and H.~M. Wiseman, ``Feedback-stabilization of an arbitrary pure state
  of a two-level atom,'' \emph{Physical Review A}, vol.~64, no.~6, pp.
  063\,810:1--9, 2001.

\bibitem{wiseman-bayesian}
H.~M. Wiseman, S.~Mancini, and J.~Wang, ``Bayesian feedback versus markovian
  feedback in a two-level atom,'' \emph{Physical Review A}, vol.~66, no.~1, pp.
  013\,807:1--9, 2002.

\bibitem{yamamoto-purestate}
N.~Yamamoto, ``Parametrization of the feedback hamiltonian realizing a pure
  steady state,'' \emph{Physical Review A}, vol.~72, no.~2, pp. 024\,104:1--4,
  2005.

\bibitem{lloyd-engineering}
S.~Lloyd and L.~Viola, ``Engineering quantum dynamics,'' \emph{Physical Review
  A}, vol.~65, no.~1, pp. 010\,101(R):1--4, 2001.

\bibitem{altafini-open}
C.~Altafini, ``Coherent control of open quantum dynamical systems,''
  \emph{Physical Review A}, vol.~70, no.~6, pp. 062\,321:1--8, 2004.

\bibitem{lendi-timedependent}
K.~Lendi, ``Extension of quantum dynamical semigroup generators for open
  systems to time-dependent hamiltonians,'' \emph{Physical Review A}, vol.~33,
  no.~5, pp. 3358--3362, 1986.

\bibitem{robin-ips}
R.~Blume-Kohout, H.~K. Ng, D.~Poulin, and L.~Viola, ``The structure of
  preserved information in quantum processes,'' \emph{Physical Review Letters},
  vol. 100, pp. 030\,501:1--4, 2008.

\bibitem{ahn-feedback}
C.~Ahn, H.~M. Wiseman, and G.~J. Milburn, ``Quantum error correction for
  continuously detected errors,'' \emph{Physical Review A}, vol.~67, no.~5, pp.
  052\,310:1--11, 2003.

\bibitem{beige}
A.~Beige, D.~Braun, B.~Tregenna, and P.~L. Knight, ``Quantum computing using
  dissipation to remain in a decoherence-free subspace,'' \emph{Physical Review
  Letters}, vol.~85, no.~8, pp. 1762--1765, 2000.

\end{thebibliography}

\end{document}